\documentclass[aps,prd,reprint,superscriptaddress,showpacs,nofootinbib]{revtex4-1}
\usepackage[final]{graphicx}
\graphicspath{{figs/}{}} 
\usepackage{amsmath,amssymb} 
\usepackage{subfigure}   

\begin{document}


\title{A Search for a Diffuse Flux of Astrophysical Muon Neutrinos with the IceCube 40-String Detector}


\affiliation{III. Physikalisches Institut, RWTH Aachen University, D-52056 Aachen, Germany}
\affiliation{Dept.~of Physics and Astronomy, University of Alabama, Tuscaloosa, AL 35487, USA}
\affiliation{Dept.~of Physics and Astronomy, University of Alaska Anchorage, 3211 Providence Dr., Anchorage, AK 99508, USA}
\affiliation{CTSPS, Clark-Atlanta University, Atlanta, GA 30314, USA}
\affiliation{School of Physics and Center for Relativistic Astrophysics, Georgia Institute of Technology, Atlanta, GA 30332, USA}
\affiliation{Dept.~of Physics, Southern University, Baton Rouge, LA 70813, USA}
\affiliation{Dept.~of Physics, University of California, Berkeley, CA 94720, USA}
\affiliation{Lawrence Berkeley National Laboratory, Berkeley, CA 94720, USA}
\affiliation{Institut f\"ur Physik, Humboldt-Universit\"at zu Berlin, D-12489 Berlin, Germany}
\affiliation{Fakult\"at f\"ur Physik \& Astronomie, Ruhr-Universit\"at Bochum, D-44780 Bochum, Germany}
\affiliation{Physikalisches Institut, Universit\"at Bonn, Nussallee 12, D-53115 Bonn, Germany}
\affiliation{Dept.~of Physics, University of the West Indies, Cave Hill Campus, Bridgetown BB11000, Barbados}
\affiliation{Universit\'e Libre de Bruxelles, Science Faculty CP230, B-1050 Brussels, Belgium}
\affiliation{Vrije Universiteit Brussel, Dienst ELEM, B-1050 Brussels, Belgium}
\affiliation{Dept.~of Physics, Chiba University, Chiba 263-8522, Japan}
\affiliation{Dept.~of Physics and Astronomy, University of Canterbury, Private Bag 4800, Christchurch, New Zealand}
\affiliation{Dept.~of Physics, University of Maryland, College Park, MD 20742, USA}
\affiliation{Dept.~of Physics and Center for Cosmology and Astro-Particle Physics, Ohio State University, Columbus, OH 43210, USA}
\affiliation{Dept.~of Astronomy, Ohio State University, Columbus, OH 43210, USA}
\affiliation{Dept.~of Physics, TU Dortmund University, D-44221 Dortmund, Germany}
\affiliation{Dept.~of Physics, University of Alberta, Edmonton, Alberta, Canada T6G 2G7}
\affiliation{Dept.~of Physics and Astronomy, University of Gent, B-9000 Gent, Belgium}
\affiliation{Max-Planck-Institut f\"ur Kernphysik, D-69177 Heidelberg, Germany}
\affiliation{Dept.~of Physics and Astronomy, University of California, Irvine, CA 92697, USA}
\affiliation{Laboratory for High Energy Physics, \'Ecole Polytechnique F\'ed\'erale, CH-1015 Lausanne, Switzerland}
\affiliation{Dept.~of Physics and Astronomy, University of Kansas, Lawrence, KS 66045, USA}
\affiliation{Dept.~of Astronomy, University of Wisconsin, Madison, WI 53706, USA}
\affiliation{Dept.~of Physics, University of Wisconsin, Madison, WI 53706, USA}
\affiliation{Institute of Physics, University of Mainz, Staudinger Weg 7, D-55099 Mainz, Germany}
\affiliation{Universit\'e de Mons, 7000 Mons, Belgium}
\affiliation{Bartol Research Institute and Department of Physics and Astronomy, University of Delaware, Newark, DE 19716, USA}
\affiliation{Dept.~of Physics, University of Oxford, 1 Keble Road, Oxford OX1 3NP, UK}
\affiliation{Dept.~of Physics, University of Wisconsin, River Falls, WI 54022, USA}
\affiliation{Oskar Klein Centre and Dept.~of Physics, Stockholm University, SE-10691 Stockholm, Sweden}
\affiliation{Dept.~of Astronomy and Astrophysics, Pennsylvania State University, University Park, PA 16802, USA}
\affiliation{Dept.~of Physics, Pennsylvania State University, University Park, PA 16802, USA}
\affiliation{Dept.~of Physics and Astronomy, Uppsala University, Box 516, S-75120 Uppsala, Sweden}
\affiliation{Dept.~of Physics, University of Wuppertal, D-42119 Wuppertal, Germany}
\affiliation{DESY, D-15735 Zeuthen, Germany}

\author{R.~Abbasi}
\affiliation{Dept.~of Physics, University of Wisconsin, Madison, WI 53706, USA}
\author{Y.~Abdou}
\affiliation{Dept.~of Physics and Astronomy, University of Gent, B-9000 Gent, Belgium}
\author{T.~Abu-Zayyad}
\affiliation{Dept.~of Physics, University of Wisconsin, River Falls, WI 54022, USA}
\author{J.~Adams}
\affiliation{Dept.~of Physics and Astronomy, University of Canterbury, Private Bag 4800, Christchurch, New Zealand}
\author{J.~A.~Aguilar}
\affiliation{Dept.~of Physics, University of Wisconsin, Madison, WI 53706, USA}
\author{M.~Ahlers}
\affiliation{Dept.~of Physics, University of Oxford, 1 Keble Road, Oxford OX1 3NP, UK}
\author{D.~Altmann}
\affiliation{III. Physikalisches Institut, RWTH Aachen University, D-52056 Aachen, Germany}
\author{K.~Andeen}
\affiliation{Dept.~of Physics, University of Wisconsin, Madison, WI 53706, USA}
\author{J.~Auffenberg}
\affiliation{Dept.~of Physics, University of Wuppertal, D-42119 Wuppertal, Germany}
\author{X.~Bai}
\affiliation{Bartol Research Institute and Department of Physics and Astronomy, University of Delaware, Newark, DE 19716, USA}
% ...

\collaboration{IceCube Collaboration}
\noaffiliation

\date{\today}

\begin{abstract}
The IceCube Neutrino Observatory is a 1 km$^{3}$ detector currently taking data at the South Pole.  One of the main strategies used to look for astrophysical neutrinos with IceCube is the search for a diffuse flux of high-energy neutrinos from unresolved sources.  A hard energy spectrum of neutrinos from isotropically distributed astrophysical sources could manifest itself as a detectable signal that may be differentiated from the atmospheric neutrino background by spectral measurement.   This analysis uses data from the IceCube detector collected in its half completed configuration which operated between April 2008 and May 2009 to search for a diffuse flux of astrophysical muon neutrinos.  A total of 12,877 upward going candidate neutrino events have been selected for this analysis. No evidence for a diffuse flux of astrophysical muon neutrinos was found in the data set leading to a 90 percent C.L. upper limit on the normalization of an $E^{-2}$ astrophysical $\nu_{\mu}$ flux of $8.9 \times 10^{-9} \ \mathrm{GeV \ cm^{-2} \ s^{-1} \ sr^{-1}}$. The analysis is sensitive in the energy range between $35 \ \mathrm{TeV} - 7 \ \mathrm{PeV}$.  The 12,877 candidate neutrino events are consistent with atmospheric muon neutrinos measured from 332 GeV to 84 TeV and no evidence for a prompt component to the atmospheric neutrino spectrum is found.  

\end{abstract}


\maketitle

\section{Introduction \label{intro}}

There are many objects in our universe that involve extremely high energy processes such as matter accreting into black holes at the centers of active galaxies and violent explosions such as  supernovae and gamma-ray bursts.  Understanding the physics of these astrophysical objects requires the observation of non-thermal high energy radiation in the form of charged cosmic rays (protons and nuclei), gamma-rays, and neutrinos. Despite progress in cosmic-ray and gamma-ray astrophysics, the nature of high energy astrophysical sources is still far from understood.  Neutrinos may elucidate the fundamental connection between the sources of high energy cosmic rays and gamma rays. 

Cosmic rays have been well studied by both space and ground based instruments.  As astrophysical messengers, their main disadvantage is that they are charged particles and thus are deflected by magnetic fields, subsequently losing their directional information.  High-energy gamma rays have been detected from many galactic and extragalactic objects, but their effectiveness as cosmic messengers over long distance scales is limited by absorption on extragalactic background light. Neutrinos could provide a fundamental connection between cosmic rays and gamma rays. 

Even if individual astrophysical neutrino sources are too weak to be detected, a superposition of all the sources may give rise to a detectable extraterrestrial flux.  In this paper, we present results from a search for a diffuse flux of astrophysical muon neutrinos performed with the IceCube detector using data collected in its half completed configuration between April 2008 and May 2009.  We first summarize astrophysical and atmospheric neutrino models in Section \ref{astronu} and describe the IceCube detector in Section \ref{icecube}. We outline in Section \ref{reconstruction} how our final neutrino sample was obtained.  The analysis methodology is discussed in detail in Section \ref{analysis} and we present our final results in Section \ref{results}.  



\section{Astrophysical and Atmospheric Neutrino Fluxes \label{astronu}}

The benchmark diffuse astrophysical $\nu_{\mu}$ search presented in this paper assumes an astrophysical flux, $\Phi$, with a spectrum $\Phi\propto E^{-2}$ resulting from shock acceleration.  In addition to the $E^{-2}$ spectral shape,  astrophysical models of varying normalization and spectral shapes were tested as well.  The Waxman-Bahcall upper bound \cite{wbbound} was derived for optically thin sources assuming a $\Phi\propto E^{-2}$ primary cosmic ray spectrum.  Becker, Biermann, and Rhode \cite{beckerfsrq} calculated the diffuse astrophysical neutrino flux from active galactic nuclei using observations from Fanaroff and Riley Class II (FR-II) radio galaxies.  These sources were used to normalize the flux of neutrinos by assuming a relationship between the disk luminosity, the luminosity in the observed radio band, and the calculated neutrino flux.  Mannheim \cite{mannheimagn} and Stecker \cite{steckeragn} derived models for optically thick Active Galactic Nuclei sources assuming the sub-TeV diffuse gamma-ray flux observed by the Compton Gamma Ray Observatory \cite{egretpaper} is produced by the decay of neutral pions. BL Lacertae objects that emit TeV gamma rays can be interpreted to be optically thin to photon-neutron interactions.   The model calculated by M\"{u}cke \textit{et al.} \cite{hbllac} assumes that charged cosmic rays are produced in these sources through the decay of escaping neutrons.  An average spectrum of neutrinos from the precursor and prompt phases of gamma-ray bursts is calculated in Ref.~\cite{grbnu} by correlating the gamma-ray emission to the observed flux of ultra high energy cosmic rays.

The primary backgrounds in the search for diffuse astrophysical $\nu_{\mu}$ are the atmospheric muons and neutrinos arising from cosmic ray induced extensive air showers.  The substantial downward-going atmospheric muon background persists over a wide energy range from primary cosmic ray energies of around a GeV to the highest measured extensive air showers of 100 EeV \cite{gaisserbook}. These events were removed by using the Earth as a filter in order to select upward-going neutrinos traversing through the Earth.  Two classes of atmospheric neutrinos were considered: neutrinos arising from the decay of pions and kaons (the conventional atmospheric neutrino flux) and neutrinos arising from the decay of charm-containing mesons (the prompt atmospheric neutrino flux). Detailed three-dimensional calculations of the energy spectrum and angular distribution of the conventional atmospheric neutrino flux are summarized in Refs. \cite{bartol} and \cite{honda:2006}.  The conventional atmospheric neutrino spectrum approximately follows an $E^{-3.7}$ spectrum in the TeV energy range. The prompt component of the atmospheric neutrino flux is yet to be measured, but full calculations of the prompt flux are given in Refs. \cite{sarcevicstd, naumov,Martin}.  The prompt component of the atmospheric neutrino flux is predicted to follow the primary cosmic ray energy spectrum which is approximately $E^{-2.7}$.  Since a hypothetical diffuse astrophysical neutrino flux would have a harder energy spectrum than atmospheric neutrino backgrounds, evidence for a diffuse flux would appear as a hardening of an energy-related observable distribution. 


\section{The IceCube Detector \label{icecube}}

\begin{figure}[htp]
\includegraphics[width=0.5\textwidth]{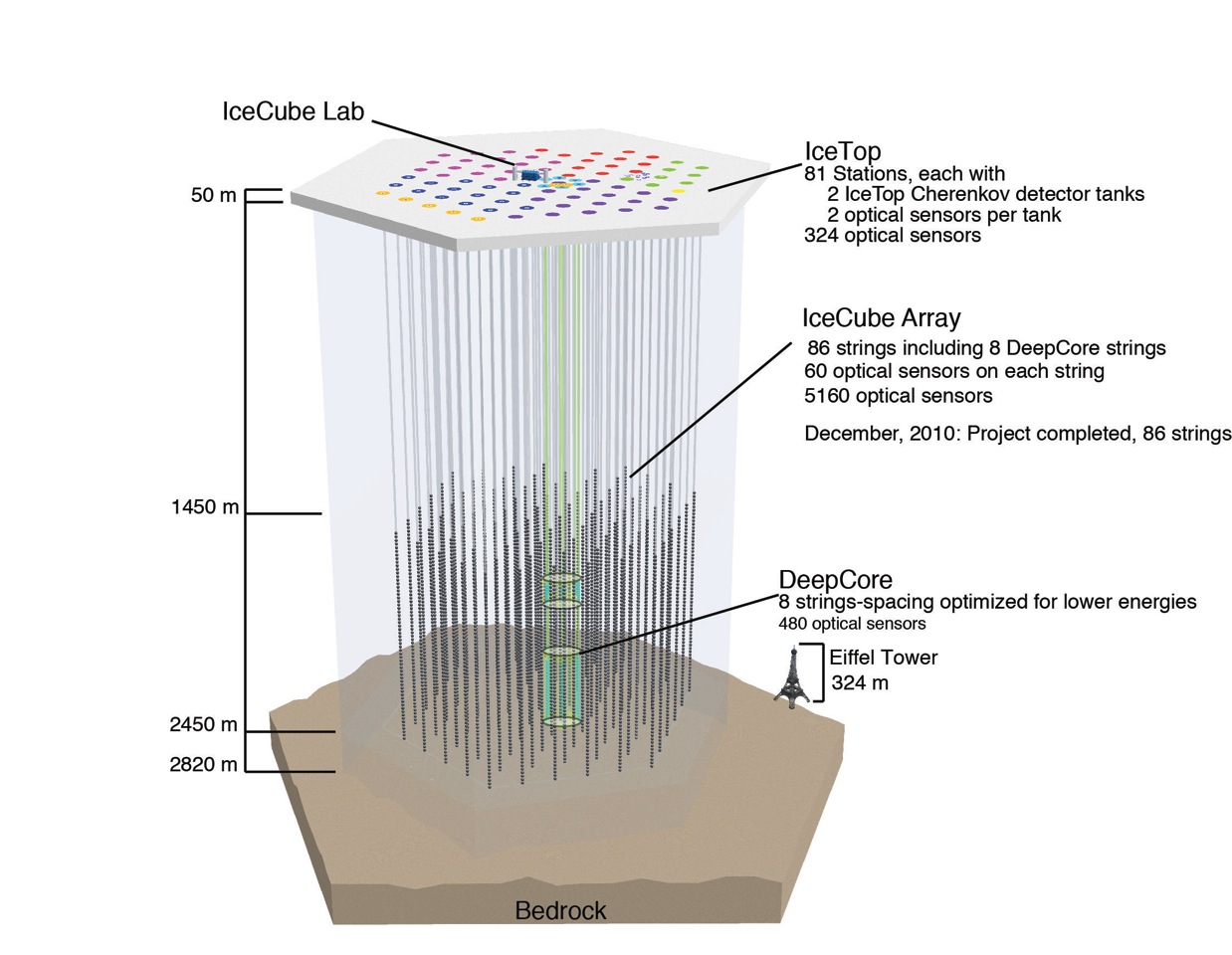}
\caption[IceCube 40-String Layout]{Three-dimensional view of the IceCube detector layout.  This work was based on the 40-string configuration which was half of the completed detector.  Its footprint is indicated by the green, red, and pink circles . The 40-string configuration was operational from April 2008 to May 2009.}  
\label{ic40layout}
\end{figure}

IceCube consists of three detectors operating together. The main in-ice array is composed of $4800$ Digital Optical Modules (DOMs) arranged in 80 strings which are deployed vertically with 60 DOMs per string.  The detector is deployed deep in the Antarctic ice between a depth of $1450$ and $2450$ meters.  The vertical spacing between each DOM is $17 \ \mathrm{m}$ and the horizontal spacing between each string of DOMs is $125$ m giving a total instrumented volume of $1 \ \mathrm{km^{3}}$.  The design is optimized for the energy range of $100 \ \mathrm{GeV}$ to $100 \ \mathrm{PeV}$ \cite{performancepaper:2006}.  The DeepCore extension is deployed within the main in-ice array and consists of six specialized strings which lower the energy reach to $10 \ \mathrm{GeV}$.  IceCube was deployed in stages with the first string deployed during the 2005-2006 Austral summer.  This analysis is based on one year of data taken with the $40$-string configuration (Fig. \ref{ic40layout}) which was deployed during the 2007-2008 Austral summer and was operational from April 2008 to May 2009.

Each DOM consists of a 13 inch (33.02 cm) pressurized sphere, a 10-inch (25 cm) Hamamatsu photomultiplier tube \cite{pmtpaper} (model R7081-02), a mu-metal magnetic shield, and associated electronics responsible for the operation and control of the PMT as well as amplification, filtering, and calibration \cite{performancepaper:2006}.  The DOMs are triggered by Cherenkov photons produced by charged particles in the Antarctic ice.  In particular, $\nu_{\mu}$-induced charged-current interactions produce muons that can traverse the entire IceCube array.  Analog waveforms captured by the PMTs are digitized in situ by the DOM main board.  The capture process is initiated by a signal derived from a discriminator connected to a high-gain signal path if the threshold (0.25 photoelectrons) is surpassed \cite{icecubedaq}.  For the data set considered in this work, the triggered event was sent to a buffer for further filtering if it satisfied a simple majority trigger (SMT) of eight triggered DOMs within a 5 $\mathrm{\mu s}$ time window.  

Below a depth of 1450 m, the Antarctic ice is free of air bubbles and exhibits exceptional optical clarity with absorption lengths ranging from 100 m to 200 m and effective scattering lengths ranging from 20 m to 70 m \cite{icepaper:2006,spicepaper}.  The scattering and absorption lengths vary due to the concentration of dust in the glacial ice, which varies quite strongly with depth \cite{spglaciology} due to varying atmospheric conditions and volcanic activity during the glacial history of Antarctica. The depth and wavelength dependence of the scattering and absorption have been measured with a variety of in-situ light sources \cite{icepaper:2006}.  The ice properties have recently been measured over the full depth range \cite{spicepaper} of the IceCube detector using the in-situ LEDs present in every DOM main board resulting in what is called the South Pole Ice (SPICE) model.  

\section{Simulation and Data Filtering \label{reconstruction}}
\subsection{Simulation}

This work required an accurate Monte Carlo simulation of the down-going atmospheric muon background,  the atmospheric neutrino flux and the subsequent detector response. The simulation was used to determine event selection criteria in order to remove the mis-reconstructed atmospheric muon background and in the profile construction method (Section \ref{analysis}) to compare the predicted neutrino energy-correlated observable distribution with the data to search for evidence of astrophysical neutrinos.  

The generation of extensive air showers initiated by high energy cosmic ray particles and the propagation of the subsequent muons through the atmosphere was handled by the CORSIKA (COsmic Ray SImulations for KAscade) \cite{corsika} event generator.  Hadronic interactions of the cosmic-ray primaries in the atmosphere were modeled using the SIBYLL \cite{sybill} interaction model.  The composition of the primary cosmic-ray spectrum was taken from the H\"{o}randel poly-gonato \cite{polygonato} model which modeled the primary cosmic ray spectrum  as a combination of two power laws for each primary particle type. 


The generation of neutrinos of all flavors was handled by the ANIS (All Neutrino Interaction Simulation) code \cite{anis}.  ANIS uses the parton structure functions from CTEQ-5 \cite{cteq5}.   Neutrinos were generated on a random position on the Earth's surface and then propagated through the Earth. The structure of the Earth is modeled by the PREM, or Preliminary Reference Earth Model \cite{prem}.  In order to reduce computation time, neutrinos that reach the detector were forced to interact with the nearby Antarctic ice or bedrock to produce secondary particles that automatically trigger the detector.  Each event was assigned a weight that represents the probability that this particular neutrino interaction occurred.  Neutrinos were typically generated with a baseline energy spectrum of either $E^{-1}$ or $E^{-2}$.  The event weights that were calculated can be used to re-weight the baseline generated spectra to any astrophysical or atmospheric neutrino model.  

A daughter muon from a muon neutrino charged current interaction or an atmospheric muon passing from the atmosphere into earth rock was propagated using the Muon Monte Carlo (MMC) \cite{mmc} code.  MMC incorporates the various continuous and stochastic energy loss mechanisms of ionization, bremsstrahlung, photo-nuclear interactions, and pair production.  The Cherenkov light produced by the muon and the various secondaries was then propagated from the muon track through the detector volume to the DOMs  by using one of two methods: numerical tabulation and direct tracking.  

The first method was provided by the Photonics \cite{photonicspaper} software package which incorporates numerically tabulated photon distribution results of various simulation runs with different light sources.  Photonics tables are computationally efficient and have the added benefit of allowing the full ice description to be used in the reconstruction of muon events.  The second method used direct photon tracking provided by the Photon Propagation Code (PPC) \cite{spicepaper} which allows for a more complete description of photon propagation in the Antarctic ice since every photon is individually tracked and propagated.  This work used PPC for the simulation of neutrinos and Photonics for the simulation of the background atmospheric muons.  This choice was made since the computational efficiency of Photonics is well suited to the generation of a large amount of atmospheric muon background simulation which subsequently helps to reduce the uncertainty of the estimated mis-reconstructed atmospheric muon background.  The numerical accuracy of PPC is appropriate for the generation of neutrino simulation which includes the atmospheric neutrino background.  Benchmark neutrino simulation sets generated with PPC and Photonics revealed that the largest discrepancy was a ~$30\%$ difference in the neutrino event rate near a prominent dust layer $2050$ m deep in the South Pole ice, whereas the overall neutrino event rate disagreement was ~$9\%$.  

\subsection{Event Selection}

The event selection strategy in this analysis used the Earth as a filter to remove all muons from cosmic ray induced extensive air showers and retain as many neutrino-induced muon events as possible.  The reconstructed energy spectrum of the neutrino sample that remained (roughly from the TeV to PeV energy range) was then analyzed using the method outlined in Section \ref{analysis} for evidence of astrophysical neutrinos. The IceCube 40-string data set used in this analysis yielded a total live time of $375.5$ days. 



The primary trigger for this analysis was a multiplicity condition which required eight DOMs to exceed their discriminator threshold within a 5 $\mu$s time window.  In addition, a local coincidence condition was enforced that requires the vertical neighbors of the triggered DOMs to trigger within 1 $\mu$s of each other.  The rate for this primary trigger was $\sim1$ kHz.  Since the trigger rate was dominated by atmospheric muons, the data was processed in several stages in order to remove the atmospheric muon background and retain only neutrino events at the final analysis level. First, the triggered event rate at the South Pole was reduced to 22 Hz by using an online software filter.  The arrival directions of the muon tracks in the IceCube detector were determined with a maximum likelihood reconstruction procedure.  The muon track geometry is uniquely described by an arrival direction and a vertex position along the track which result in five degrees of freedom for the reconstruction.  The likelihood function \cite{recopaper:2004} parametrizes the probability of observing the Cherenkov photon arrival times given the muon track geometry.  Preliminary reconstructions were performed using a single photoelectron (SPE) likelihood which utilizes the arrival time of the first Cherenkov photon arriving in each DOM.  All events reconstructed as upward going through the Earth ($\theta > 90^{\circ}$) were kept in the initial first stage of filtering.  Events reconstructed as down-going must pass an energy cut that tightens for more vertical events.  This ensures that truly up-going high energy events, initially reconstructed as down-going, may be correctly reconstructed and kept in the final up-going event sample. 

\begin{table}[htp]
\centering
\begin{tabular}{l}
\hline \hline
Observable and Selection Criteria \\ [0.5ex]
\hline
$\theta > 90^{\circ}$ \\
$\frac{\log(L)}{(N_{\mathrm{ch}}-5)} < 8$ OR $\frac{\log(L)}{(N_{\mathrm{ch}}-2.5)} < 7.1$ \\
$\sigma < 3^{\circ}$ \\
$\log(L_{\mathrm{Bayes}}/L) > 25$ for $\cos(\theta) < -0.2$ \\
$\log(L_{\mathrm{Bayes}}/L) > (75\cos(\theta) + 40)$ for $\cos(\theta) > -0.2$\\
$\log(\frac{L_{\mathrm{Bayes1}}+L_{\mathrm{Bayes2}}}{L}) > 35$ \\
$\theta_{\mathrm{SplitTime}} > 80^{\circ}$ \\
$\theta_{\mathrm{SplitGeo}} > 80^{\circ}$ \\
$\mathrm{NDir} > 5$ \\
$\mathrm{LDir} > 240$ \\
$|\mathrm{SDir}| < 0.52$ \\
\hline
\end{tabular}
\caption[Final Purity Cuts]{Summary of the analysis level selection criteria applied to the IceCube data, neutrino simulation, and the atmospheric muon background simulation to obtain the final event sample for the analysis.}
\label{nucuts}
\end{table}

The second filtering stage involved more CPU intensive reconstructions performed offline outside of the South Pole.  Among these reconstructions is the multiple photoelectron (MPE) fit which utilizes the likelihood description of the arrival time of the first Cherenkov photon given N expected photons.  The first photon is less scattered than the average single photon and hence the likelihood description of the detected photoelectron is modified when this information is taken into account The MPE likelihood is a more sophisticated likelihood description than the SPE likelihood reconstruction.  It gives improved direction resolution at higher energies.  Estimates of the muon energy (see the next section), the angular resolution, and quality parameters used for background rejection are calculated during the offline processing stage.  About $5\%$ of the cosmic ray-induced muons in the atmosphere that trigger the IceCube detector are mis-reconstructed as going up through the Earth and need to be separated from neutrino-induced muons at the final analysis level.  This is accomplished using quality criteria which are based on parameters derived from the reconstructed muon track. Table \ref{nucuts} summarizes the analysis cuts applied to the level 1 filtered data and simulation. Table \ref{passrate2} summarizes the number of data and simulation events that satisfied each successive analysis cut defined in Table \ref{nucuts}.  The quality parameters used to obtain the final analysis sample are:

\begin{figure*}[htp]

\includegraphics[width=0.49\textwidth]{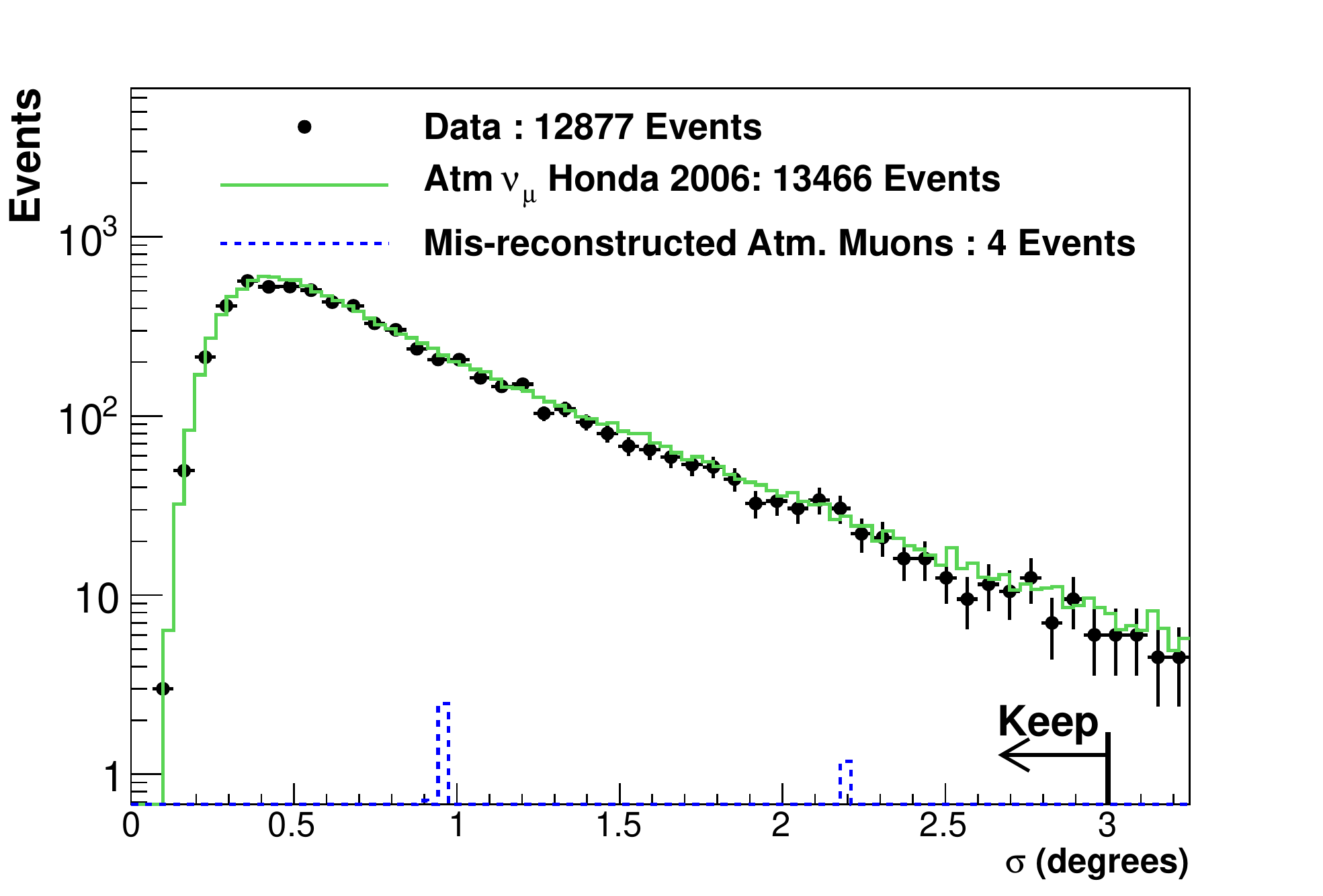} 
\includegraphics[width=0.49\textwidth]{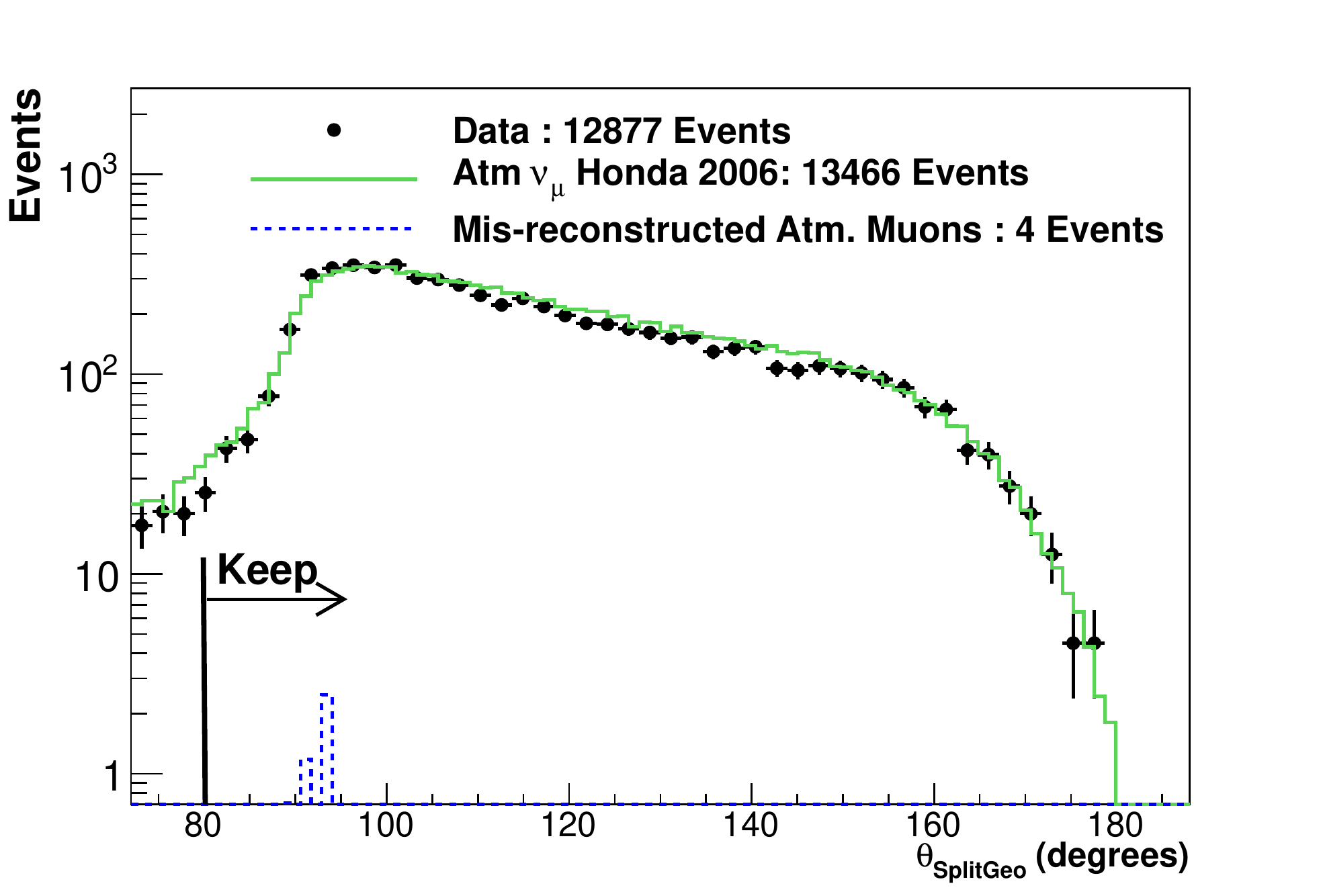} \\
\includegraphics[width=0.49\textwidth]{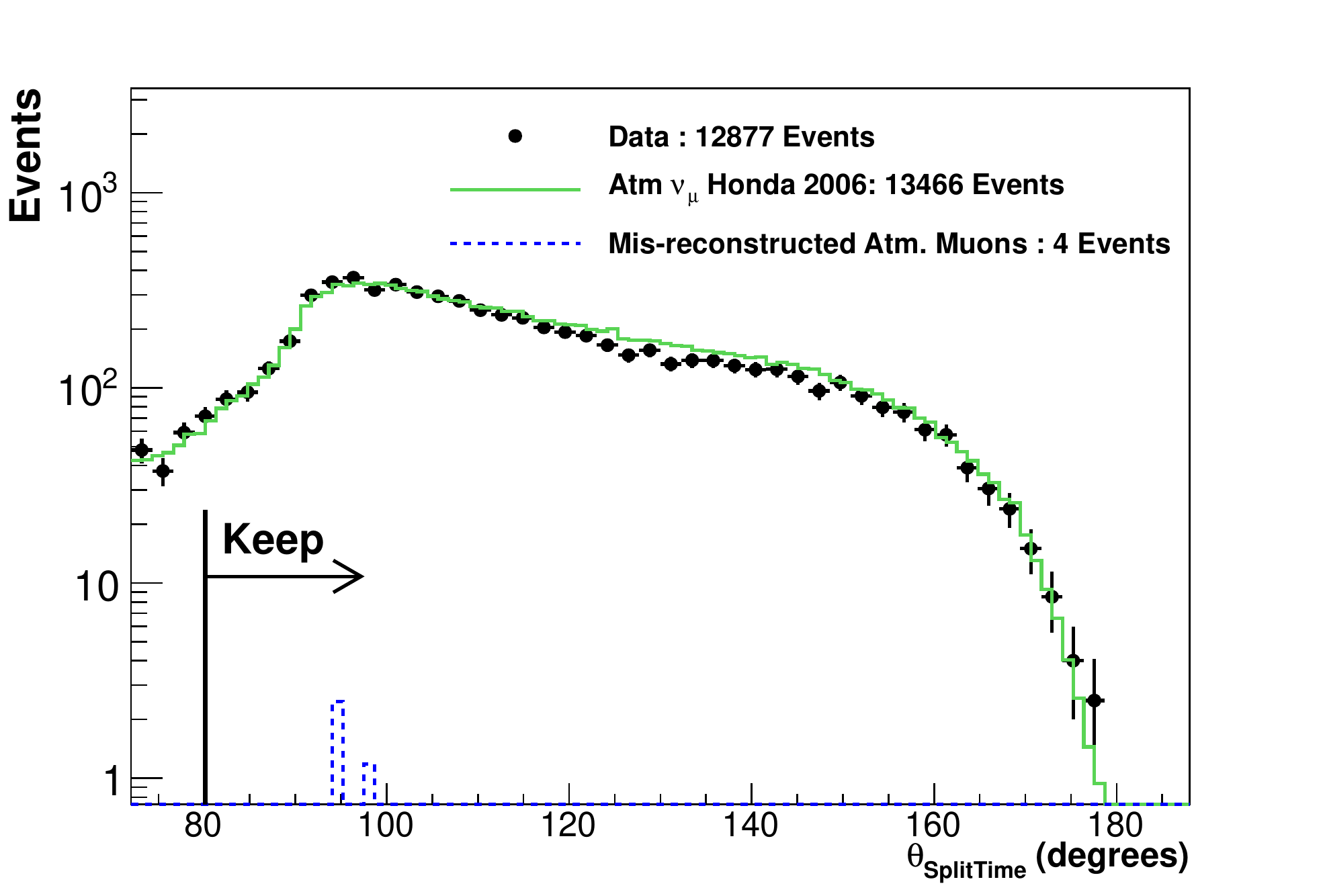} 
\includegraphics[width=0.49\textwidth]{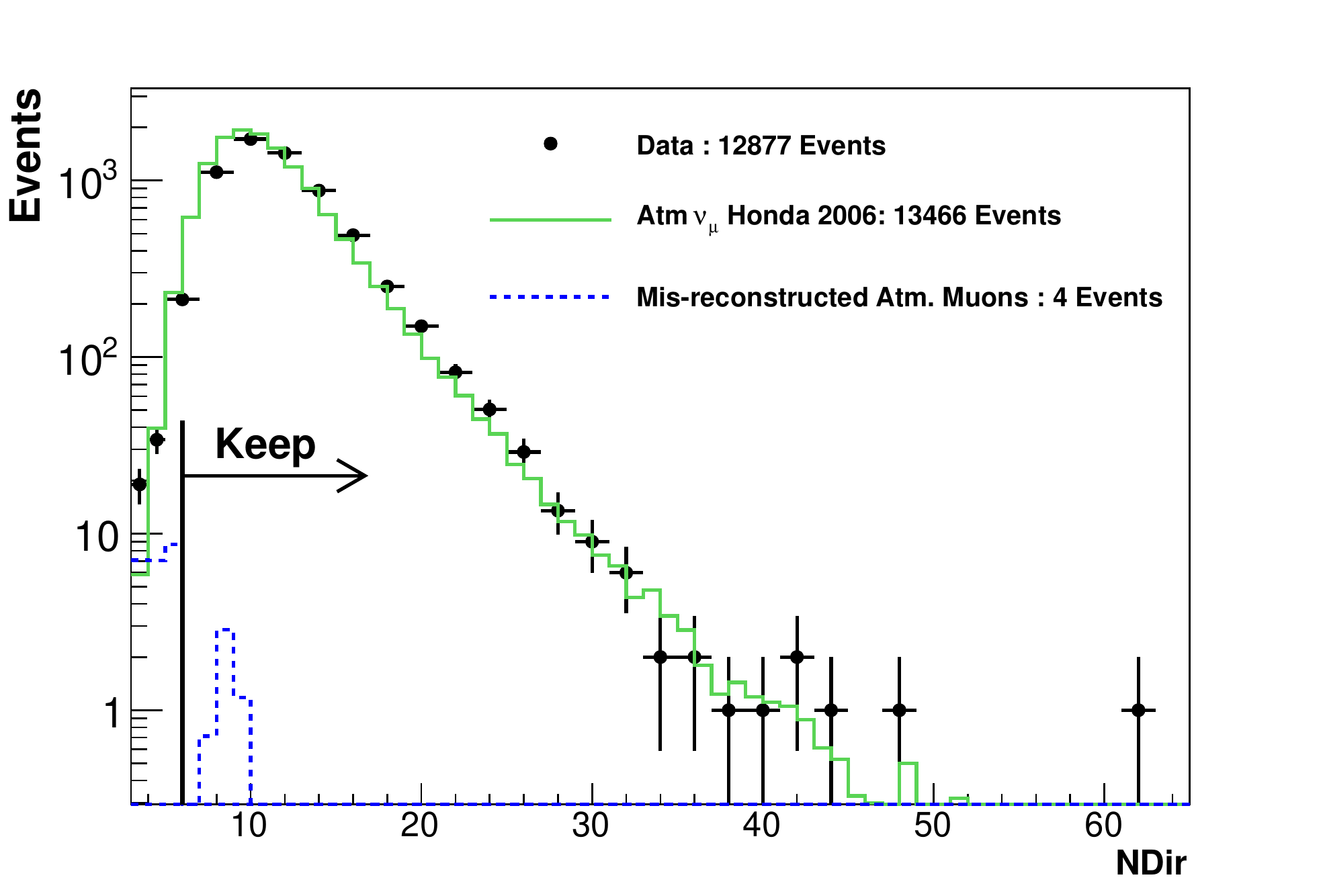} \\
\includegraphics[width=0.49\textwidth]{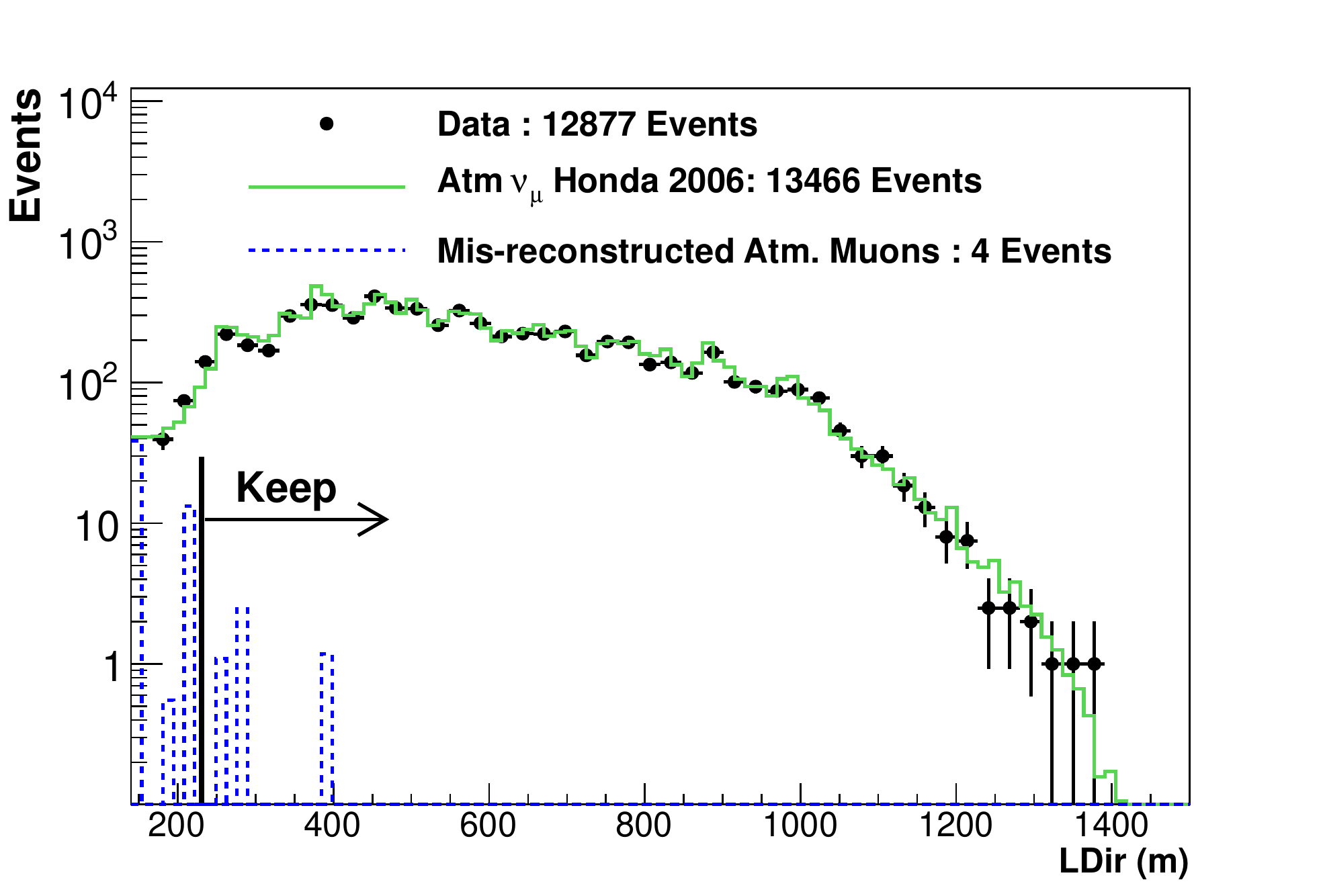} 
\includegraphics[width=0.49\textwidth]{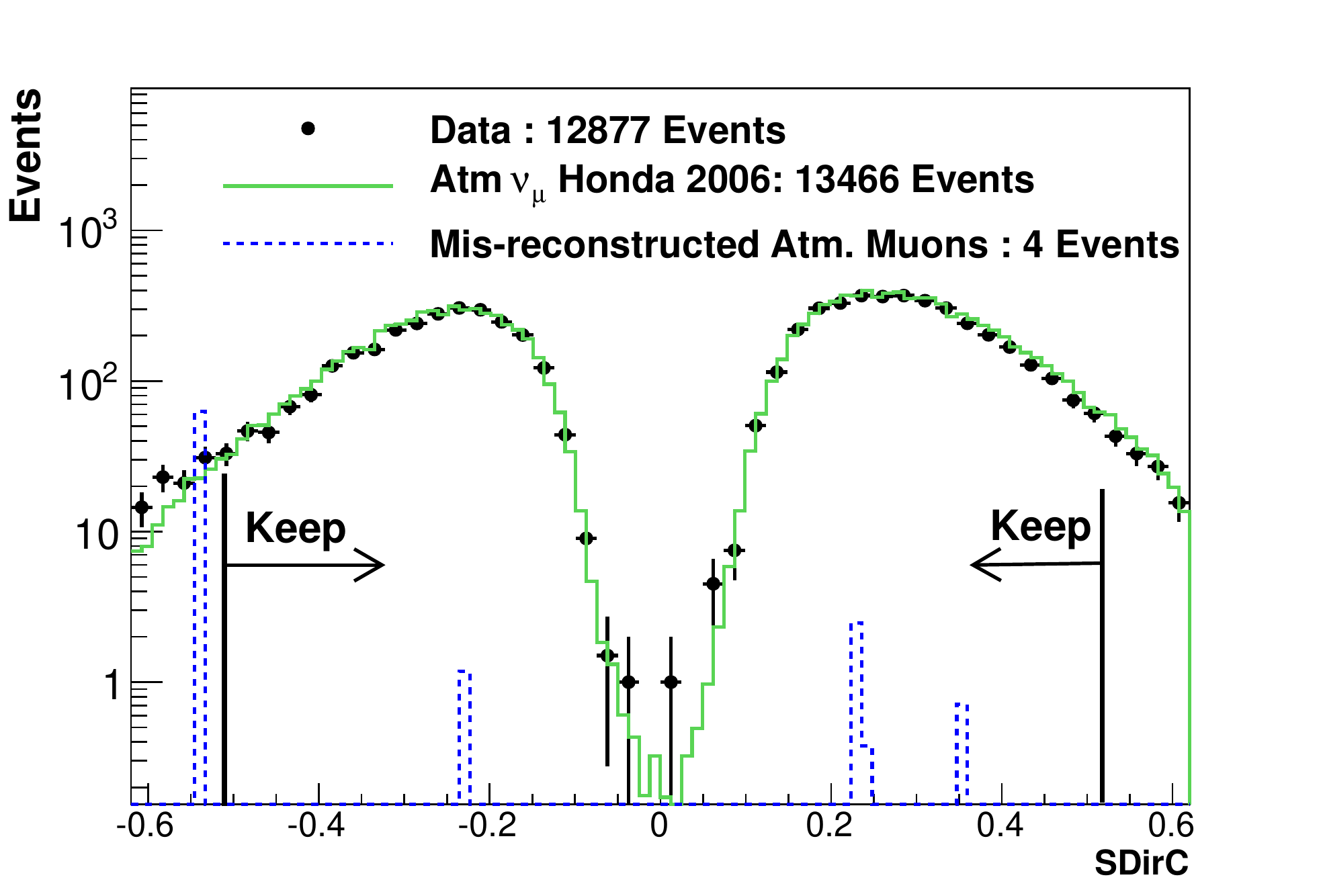} \\
\caption[Analysis Level Data/MC Comparisons for Track Quality Variables]{Track quality observables for data (black), atmospheric neutrino simulation (green) and mis-reconstructed atmospheric muon simulation (blue) after all analysis cuts have been applied.}  
\label{finalcutplots}
\end{figure*}

\begin{figure*}[htp]
\includegraphics[width=0.49\textwidth]{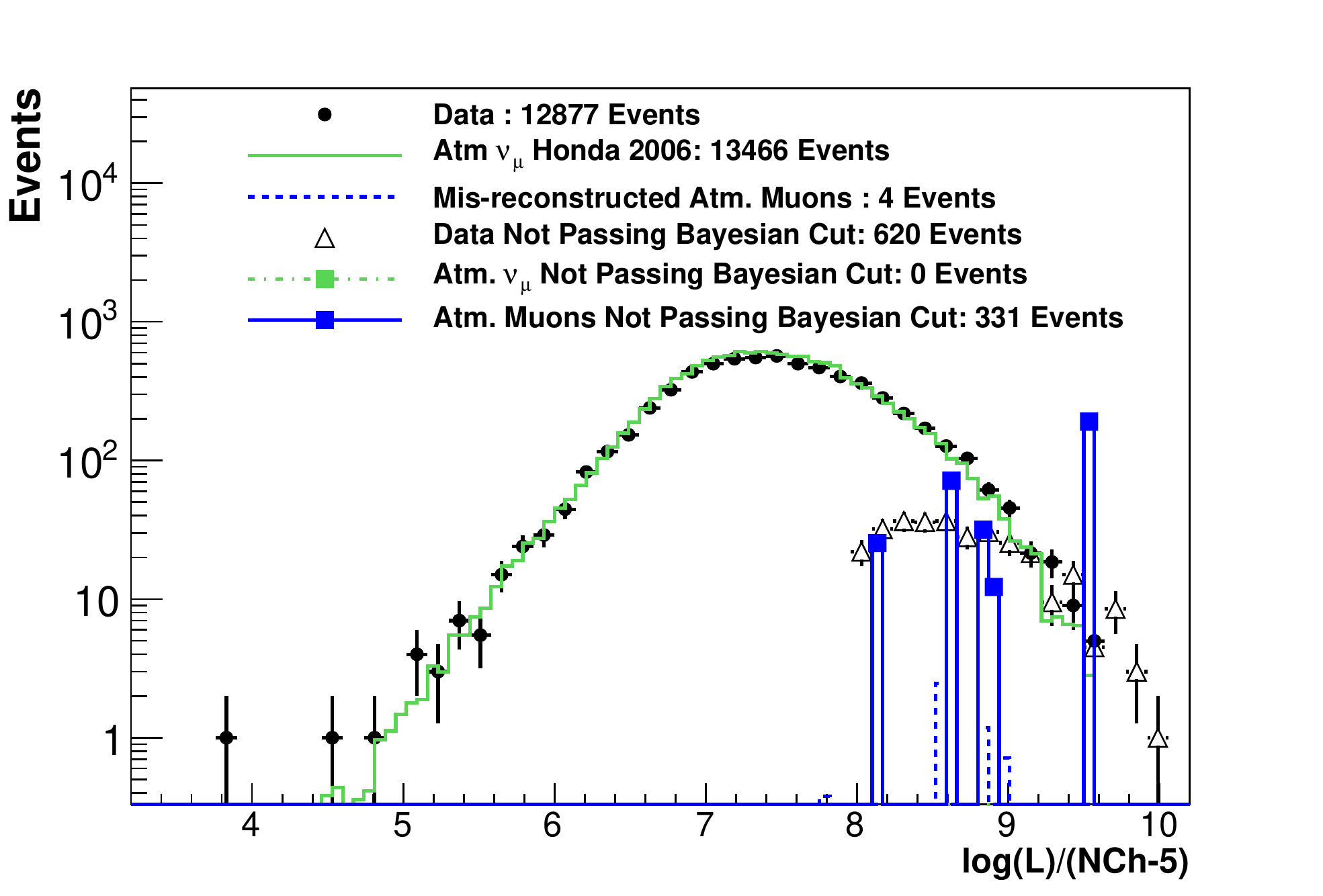}
\includegraphics[width=0.49\textwidth]{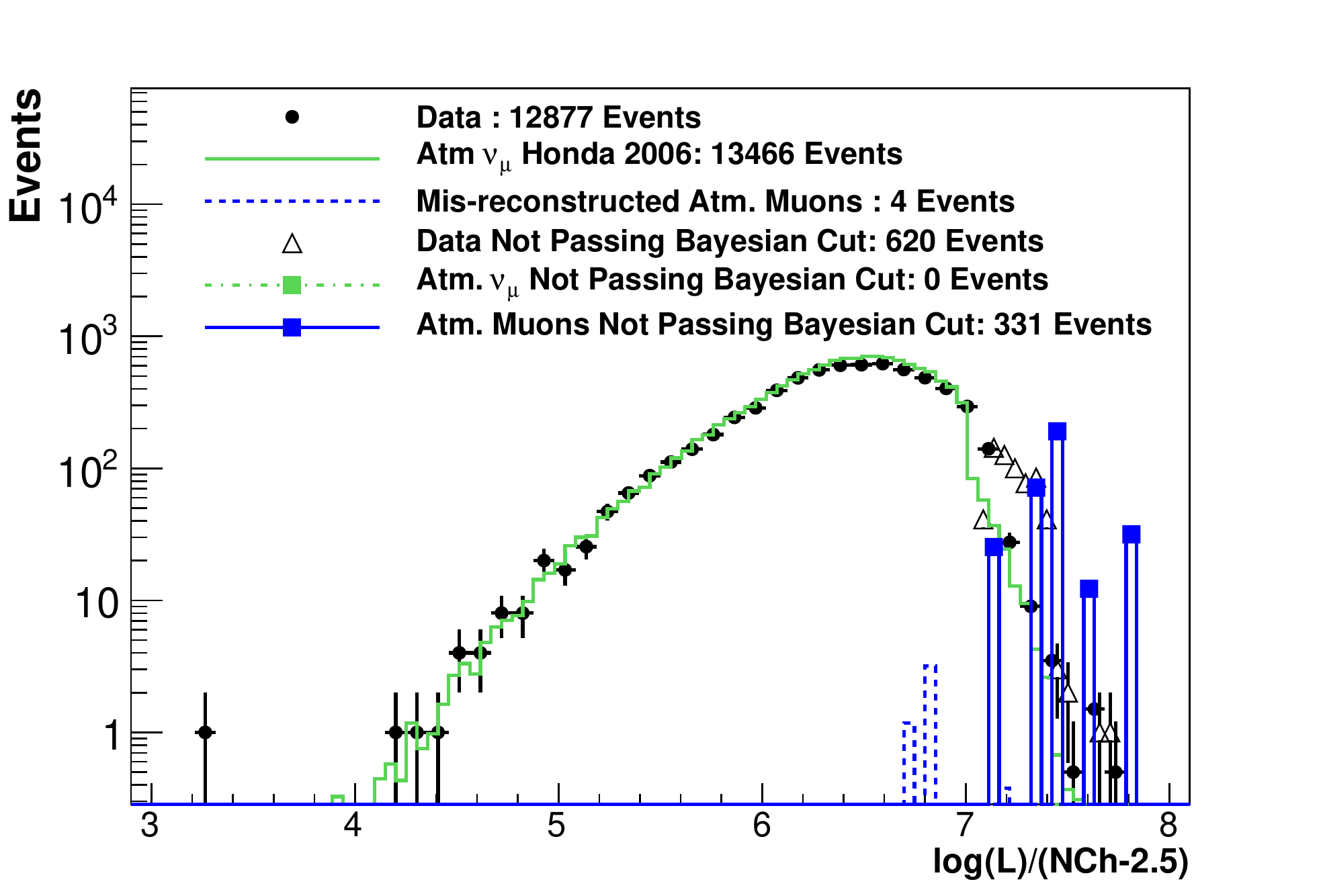} \\
\includegraphics[width=0.49\textwidth]{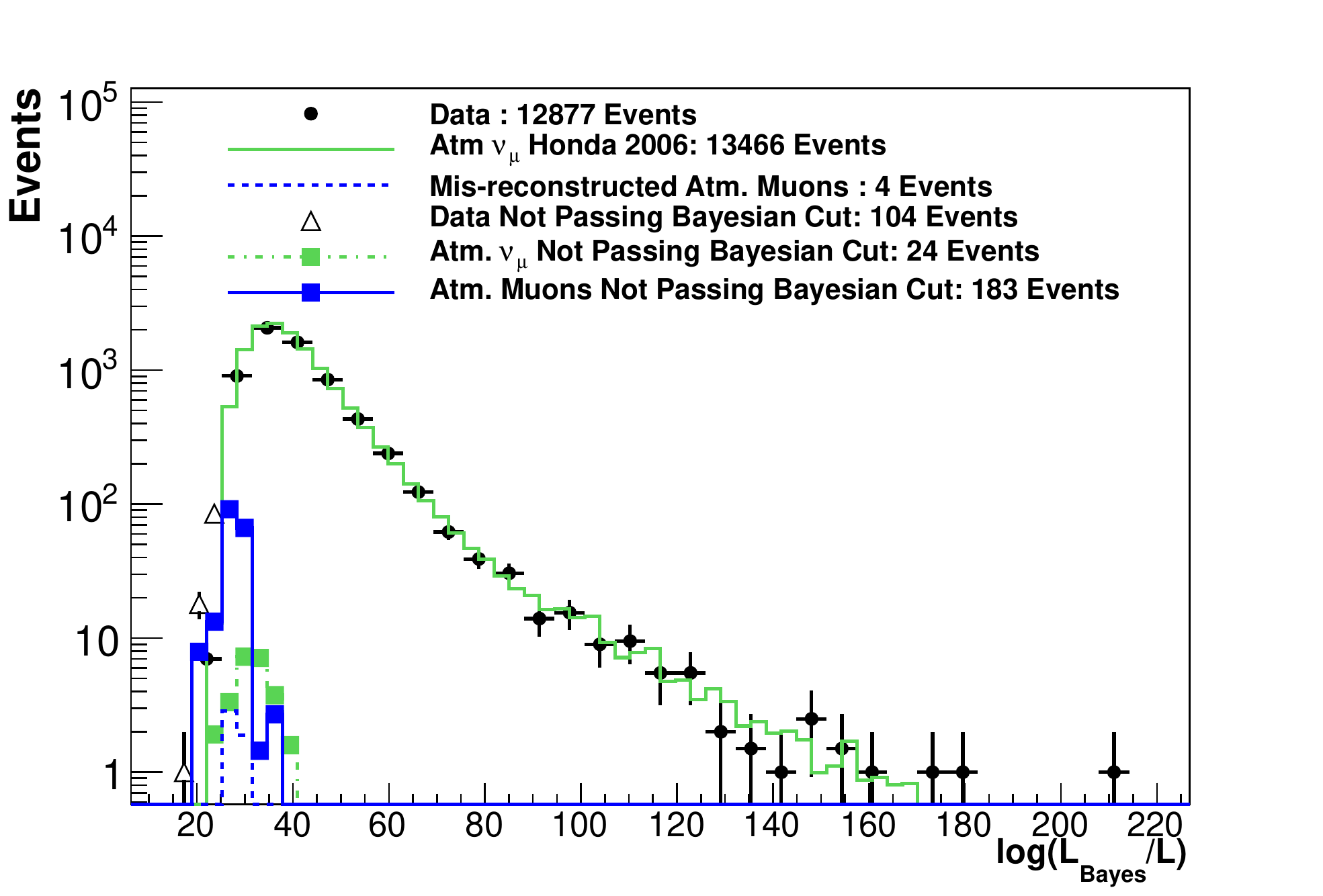} 
\includegraphics[width=0.49\textwidth]{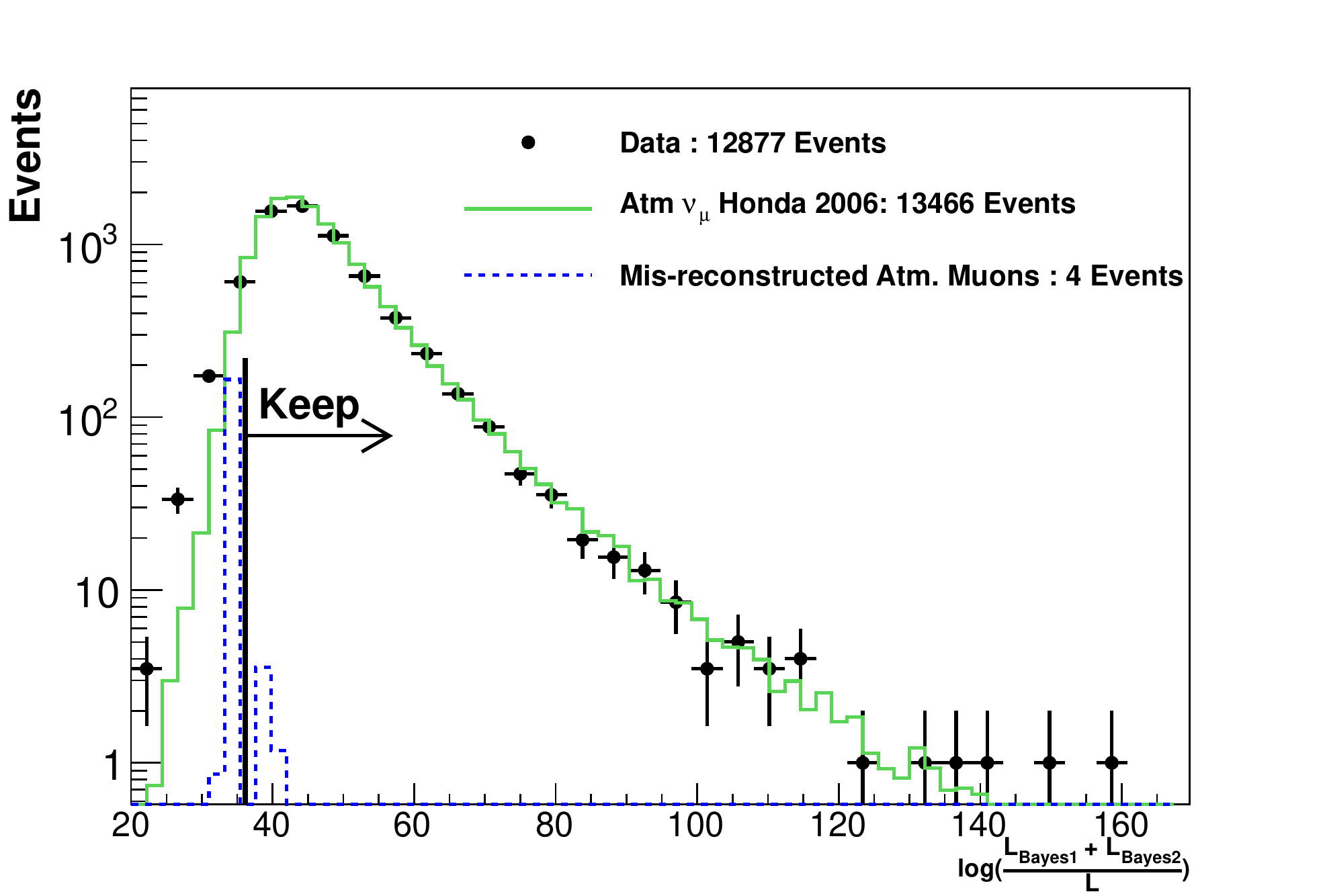} \\
\caption[Analysis Level Data/MC Comparisons for Track Quality Variables Split fits]{Reduced Log-likelihood and Bayesian likelihood ratio quality observables for data (black), atmospheric neutrino simulation (green), and atmospheric muon simulation (blue) after all analysis cuts have been applied.  Events that do not pass the final two-dimensional quality criteria for the reduced log-likelihood and the Bayesian likelihood ratio are also shown.}
\label{finalcutplotssplit}
\end{figure*}

\begin{itemize}
\item \textbf{Reconstructed zenith angle ($\theta$)}:  The zenith angle of the reconstructed muon track is used as a cut parameter to select muon events with reconstructed directions that traverse through the Earth. 

\item \textbf{Reduced log-likelihood}:  The log-likelihood value of the reconstructed track was divided by the number of degrees of freedom of the fit. The number of degrees of freedom is given by the number of triggered DOMs ($N_{\mathrm{ch}}$) minus five, which is the number of free parameters in the reconstruction.  Since $N_{\mathrm{ch}}$ loosely correlates with the muon energy, the reduced log-likelihood should be approximately energy independent. A smaller value indicates that the Cherenkov photons arrived at the individual DOMs more consistent with the likelihood description of photon arrival times.  It is an efficient observable for separating high energy atmospheric neutrinos from mis-reconstructed atmospheric muons.  This variable was found not to be energy independent for lower energy atmospheric neutrinos, however, and was subsequently found to be not efficient at background rejection at lower energies.  This was resolved empirically by redefining the effective degrees of freedom to $N_{\mathrm{ch}} - 2.5$  for low values of $N_{\mathrm{ch}}$. 
  
\item \textbf{Error estimate from the MPE reconstruction ($\sigma$)}:  The directional error ellipse for the MPE log-likelihood reconstruction was estimated following \cite{paraboloid}.   It provides an event by event $1 \sigma$ uncertainty of the arrival direction in the likelihood function used in the reconstruction of muon tracks.  

\item \textbf{Minimum zenith angle of a two-muon reconstruction ($\theta_{\mathrm{SplitGeo}},\theta_{\mathrm{SplitTime}})$:} A substantial fraction of the atmospheric muon background results from two or more muons triggering the IceCube detector during the trigger window.  In order to reduce this background, two muons were reconstructed for each event after splitting the triggered DOMs in two groups.  The separation is  accomplished one of two ways. The first uses a geometric approach by constructing a plane perpendicular to the MPE-reconstructed track while containing the average Cherenkov photon arrival positions.  The second method is performed temporally by using the mean Cherenkov photon arrival time. Each group of DOMs are used to reconstruct a single muon hypothesis resulting in two reconstructed muon tracks.  Requiring the zenith angle from both reconstructed tracks to traverse through the Earth reduces the coincident atmospheric muon background.  

\item \textbf{Log-likelihood ratio between a zenith-weighted Bayesian reconstruction and a standard reconstruction:}  The Bayesian likelihood ratio compares the hypothesis of an up-going muon track with the alternative hypothesis of a down-going muon track consistent with the known zenith-dependent flux of atmospheric muons.  The Bayesian likelihood reconstruction is performed by minimizing the product of the standard likelihood and a Bayesian prior.  The Bayesian prior is based on the known zenith dependence of the down-going muon flux.  Since the prior goes to zero near the horizon, the reconstruction always results in a down-going muon.  Low values of the negative log-likelihood ratio support the alternative hypothesis of a down-going muon, whereas higher values indicate an up-going muon track.  Further details are found in Ref. \cite{recopaper:2004}.  The likelihood ratio is zenith-dependent and our selection criterion based on this quality parameter varies with the zenith angle of the MPE reconstructed track.      

\item \textbf{Log-likelihood ratio between a zenith-weighted two-muon Bayesian reconstruction and a standard reconstruction:}  The two-muon Bayesian likelihood ratio compares the hypothesis of a single up-going muon track with the alternative hypothesis of two down-going muon tracks consistent with the known zenith-dependent flux of atmospheric muons.  Two down-going muons were reconstructed separately using the DOM splitting strategies discussed above.  Each muon is reconstructed with a Bayesian prior defined with a zenith-dependent weight of the atmospheric muon flux. This observable is constructed to reject mis-reconstructed coincident atmospheric muons.  As in the single muon case, low values support the alternative hypothesis of two down-going atmospheric muons whereas higher values indicate an up-going muon track.  
    
\item \textbf{Number of DOMs with direct photoelectrons, (NDir):}  The number of Cherenkov photons arriving between $-15$ and $+75$ ns of their expected un-scattered photon arrival times from a reconstructed track is known as the number of direct photons, or $\mathrm{NDir}$ \cite{recopaper:2004}.  More direct photons would indicate a better reconstructed track.   

\item \textbf{Direct length of the reconstructed track (LDir):}  The number of direct photons, NDir, are projected back onto the reconstructed track.  The direct length, LDir \cite{recopaper:2004}, is the maximum separation distance between these projected photons.  

\item \textbf{Smoothness of the reconstructed track (SDir)}:  The direct photons (NDir) are again projected back onto the reconstructed track.  The smoothness, SDir, is a measurement of how uniformly distributed these projected photons are along the reconstructed track.  The smoothness parameter is defined between $-1$ and $1$.  Positive values of smoothness indicate that the projected photons cluster at the beginning of the track, whereas negative values of smoothness indicate there are more at the end of the track.  A smoothness that is close to $0$ indicates a uniform distribution of projected Cherenkov photons. Further details of the smoothness parameter can be found in Ref. \cite{recopaper:2004}.
\end{itemize}

\begin{table*}[htp]
\centering
\begin{tabular}{c c c c c c}
\hline \hline
Purity Criterion& Data & Total Atm. $\mu$ & Coincident $\mu$ & Atm. $\nu_{\mu}$ & $E^{-2} \ \nu_{\mu}$ \\ [0.5ex]
\hline
Triggered & $3.3\times10^{10}$ & $2.98\times10^{10}$ & $1.72\times10^{10}$ &$1\times10^{6}$&$1.03\times10^{4}$\\ 
L1 Filter & $8.0\times10^{8}$ & $7.5\times10^{8}$ &  $3.9\times10^{8}$ &$1.14\times10^{5}$& 1,956\\ 
$\theta > 90^{\circ}$ & $2.4\times10^{8}$ & $3.0\times10^{8}$ & $1.79\times10^{8}$ & 91,246 & 1,353 \\
$\log(L)$ & $8.46\times10^{6}$ & $4.58\times10^{6}$ & $1.12\times10^{6}$ & 43,183 & 934  \\ 
$\sigma$ & $1.43\times10^{6}$ & $1.05\times10^{6}$ & $4.1\times10^{5}$ & 37,174 & 677 \\ 
$\log(L_{\mathrm{Bayes}}/L)$ & $2.88\times10^{5}$ & $2.73\times10^{5}$ & $2.36\times10^{5}$ & 27,411 & 659 \\
$\log(\frac{L_{\mathrm{Bayes1}}+L_{\mathrm{Bayes2}}}{L})$ & 44,309 & 24,032 & 17,648 & 18,400 & 622 \\
$\theta_{\mathrm{SplitTime}}$ & 22,154 & 3,004 & 2253 & 15,771 & 556 \\
$\theta_{\mathrm{SplitGeo}}$& 17,648 & 1,126 & 751 & 15,020& 532 \\
$\mathrm{NDir}$& 15,771 & 751 & 370 & 14,645& 524 \\
$\mathrm{LDir}$ & 13,518 & 374 & 325 & 14,269 & 499 \\
$\mathrm{SDir}$& 12,877 & 4 & 0 & 13,466 & 475 \\
\hline
\end{tabular}
\caption[Data and MC Passing Rates for Successive Purity Cuts]{Number of events at each purity level for data and simulation for atmospheric muons, conventional atmospheric $\nu_{\mu}$, and $E^{-2}$ astrophysical $\nu_{\mu}$ with a normalization of  $N_{a}=10^{-7} \mathrm{GeV \ cm^{-2} \ s^{-1} \ sr^{-1}}$ for the full 40-string data set of $375.5$ days. The background atmospheric $\mu$ was simulated with a total live-time of 11 days.  A weighting scheme was used to increase the live-time at high energies resulting in 240 days of background live-time above a primary cosmic ray energy of 100 TeV.   The quality parameter used for the purity cut is shown and the specific values of the cuts are defined in Table \ref{nucuts}.}
\label{passrate2}
\end{table*}



\begin{figure}[htp]
\includegraphics[width=0.52\textwidth]{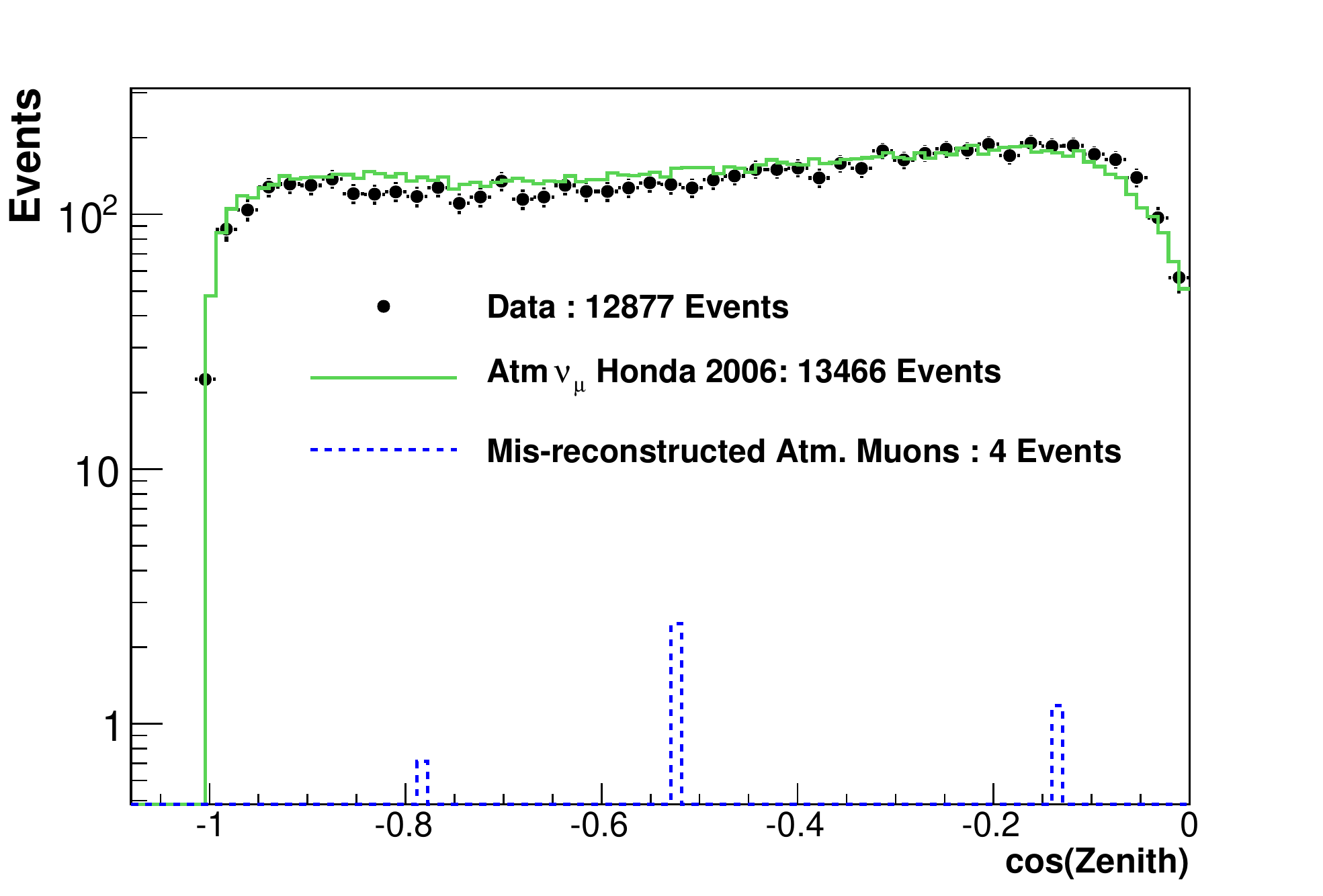} \\
\caption[Zenith Progression]{The zenith angle distribution at final analysis level for $375.5$ days of IceCube 40-string data (black) and atmospheric neutrino Monte Carlo(green).}  
\label{coszen}
\end{figure}

\begin{figure}[htp]
\includegraphics[width=0.55\textwidth]{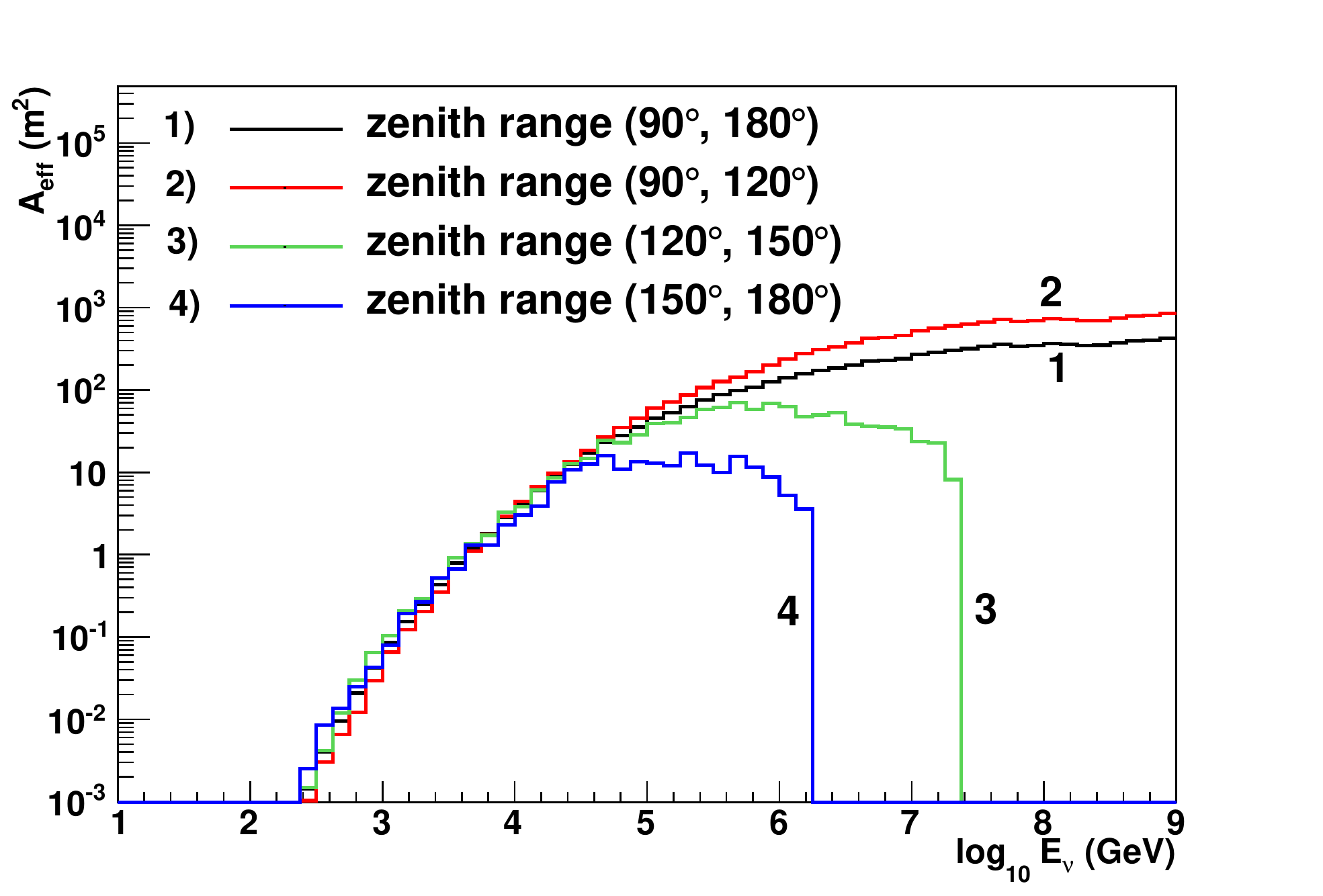}
\caption[IceCube 40-String Effective Area]{Effective area for $\nu_{\mu}+\bar{\nu}_{\mu}$ as a function of the true neutrino energy in intervals of the true zenith angle of the neutrino.  The angle averaged area is represented by the solid black line. }  
\label{effareaic40}
\end{figure}

After all analysis level cuts have been applied, we were left with $12,877$ candidate neutrino events below the horizon for the IceCube 40-string data set\footnote{\hbox{A table of the event sample used in this analysis is available at} http://www.icecube.wisc.edu/science/data/.}.  These cuts were designed in particular to maximize the retention efficiency of the simulated $E^{-2}$ astrophysical neutrino flux, which is $35.1\%$ with respect to up-going events passing the level 1 filter.  The final analysis level (after all analysis cuts have been applied) distributions for the track quality observables summarized above are shown in Figs. \ref{finalcutplots}-\ref{finalcutplotssplit} for data and Monte Carlo simulation. The zenith distribution at the final analysis level is shown in Fig. \ref{coszen}.  The background atmospheric muon contamination was estimated to be 4 events in the final sample with a relative error of $60\%$.  The background contamination was estimated from simulated down-going atmospheric muons that survived the analysis cuts. To estimate the background contamination, one would ideally have as much simulated background live-time as the data.  In practice, the simulated background live-time was significantly less than the live-time of the data with eleven simulated days over all energies.  A weighting scheme was used to increase the number of generated events at high energies resulting in 240 days of equivalent background live-time above a primary cosmic ray energy of 100 TeV.  The simulated atmospheric muons over all energy decades were then extrapolated to one year of live-time.  

The efficiency of neutrino detection for a particular analysis, which includes the efficiency of the analysis level cuts and physical effects like the absorption due to the Earth, can be characterized by the effective area which is defined as the area $A_{\textrm{eff}}(E,\theta,\phi)$ of a detector that would have a $100\%$ neutrino detection efficiency.  The total number of detected events is: \begin{equation}
N_{\mathrm{events}}=\int \mathrm{d}E_{\nu} \ \mathrm{d}\Omega \ \mathrm{d}t \ \Phi_{\nu}(E_{\nu},\theta,\phi)A_{\textrm{eff}}(E,\theta,\phi)
\label{effarea}
\end{equation} Fig. \ref{effareaic40} shows the effective area for $\nu_{\mu}+\bar{\nu}_{\mu}$ as a function of energy for this analysis averaged over different zenith angle ranges.   

\begin{figure*}[htp]
\centering
\includegraphics[width=0.50\textwidth,height=0.2625\textheight]{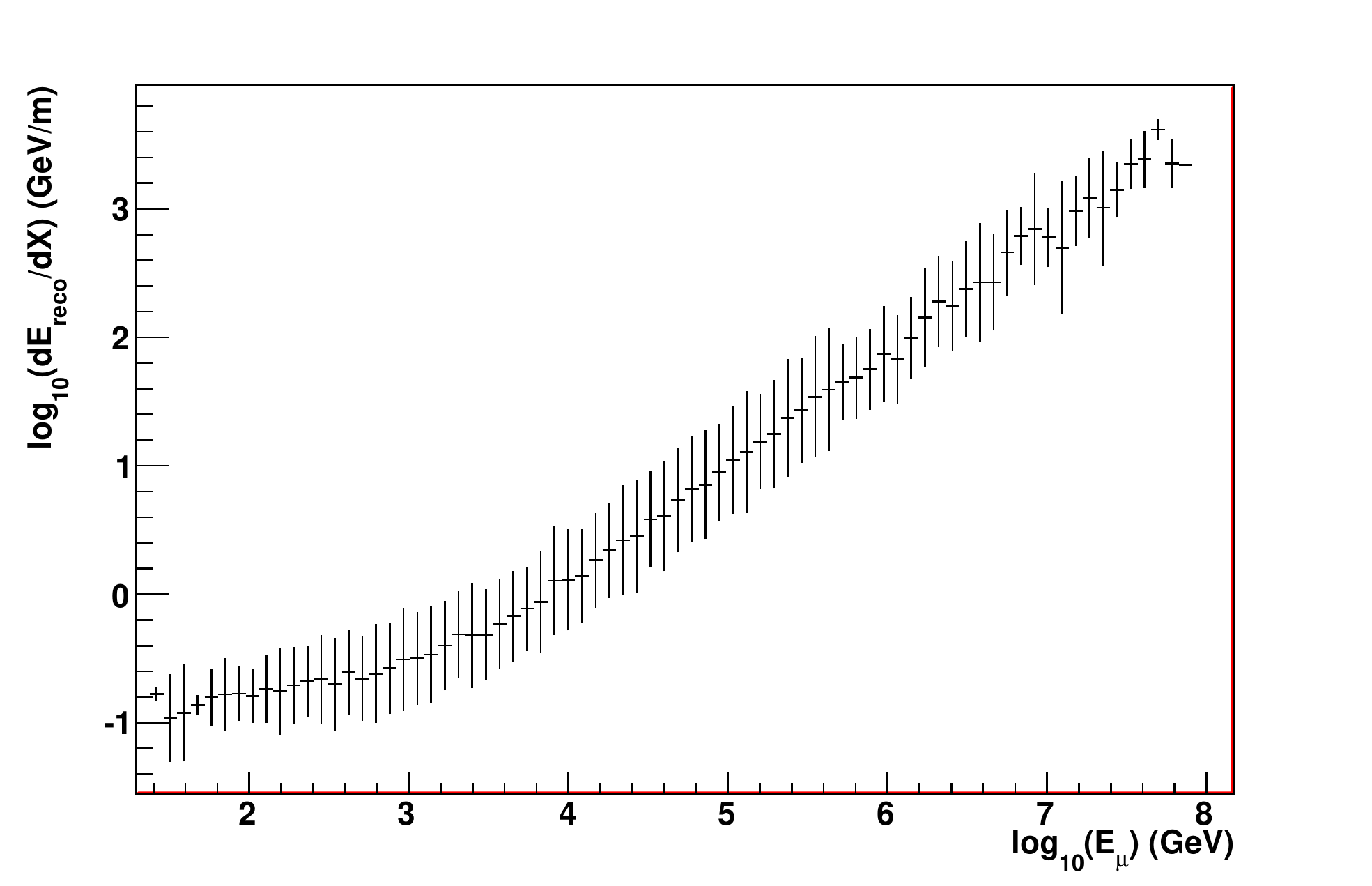}
\includegraphics[width=0.475\textwidth]{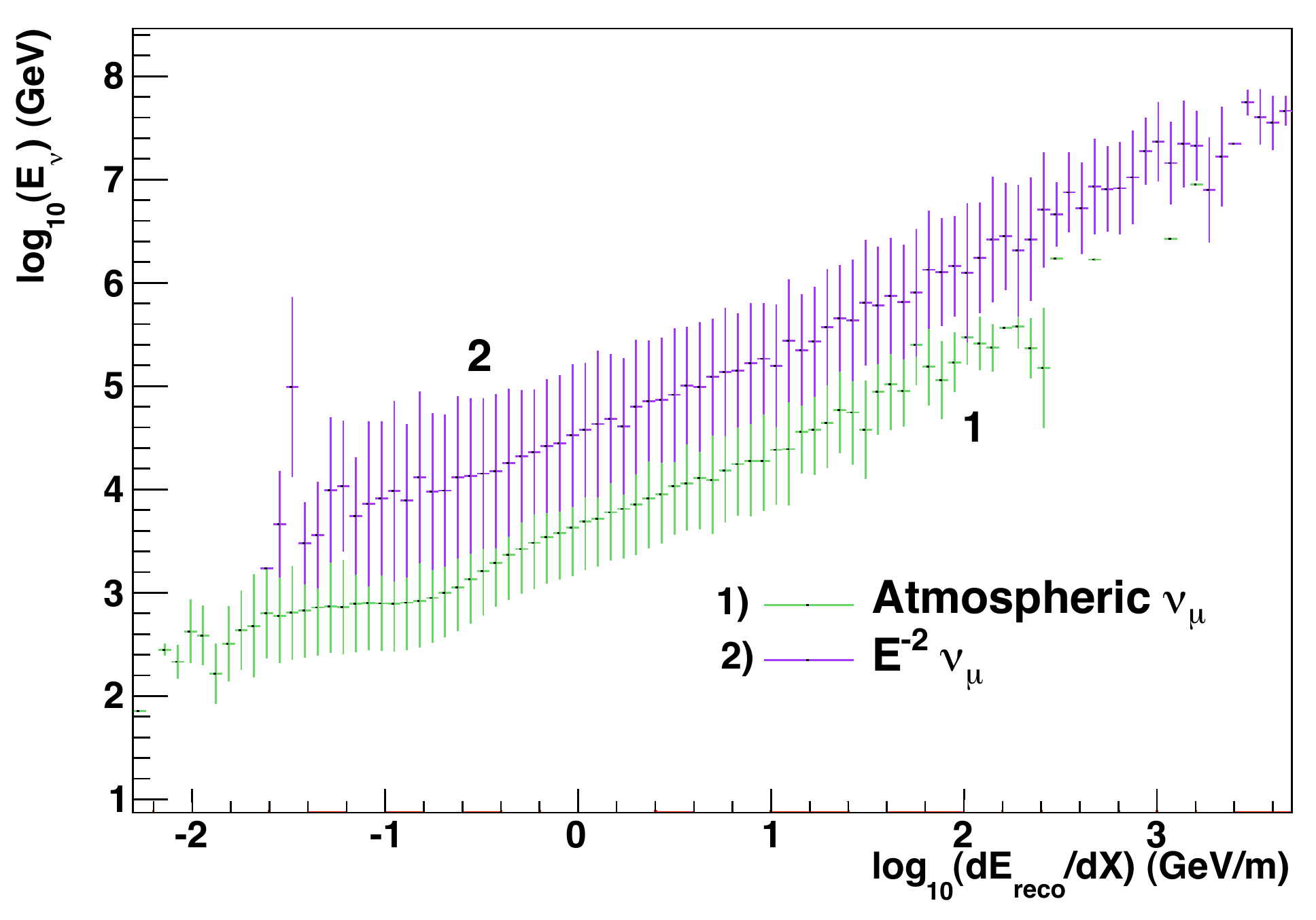}
\caption[$\mathrm{d}E_{\mathrm{reco}}/\mathrm{d}X$ Correlation with Muon Energy]{The left plot shows a profile of the average reconstructed muon energy loss $\mathrm{d}E_{\mathrm{reco}}/\mathrm{d}X$ vs. the true energy of the muon at closest approach. The error bars indicate the RMS of the reconstructed $\mathrm{d}E/\mathrm{d}X$ for a given slice in the parent muon energy. The right plot shows a profile of the energy of the primary neutrino for different bands in reconstructed $\mathrm{d}E/\mathrm{d}X$.  The error bars indicate the RMS of the parent neutrino energy.  Shown are spectra for atmospheric neutrinos and a hypothetical $E^{-2}$ astrophysical $\nu_{\mu}$ flux. }
\label{dedxscatter}
\end{figure*}

The final sample of candidate neutrino events is analyzed for the presence of astrophysical neutrinos.  The astrophysical models considered predict a flavor ratio at the source of $\nu_{\mu}:\nu_{e}:\nu_{\tau}=2:1:0$, which subsequently oscillate to a flavor ratio of $\nu_{\mu}:\nu_{e}:\nu_{\tau}=1:1:1$ at Earth.  Tau neutrinos that propagate through the Earth undergo a regeneration effect with a branching ratio $\tau \rightarrow \mu \nu_{\mu} \nu_{\tau}$ of $17\%$ and this $\nu_{\mu}$ contribution was taken into account by incorporating a separate $\nu_{\tau}$ Monte Carlo simulation. The final astrophysical $\Phi_{\mu}$ results were derived assuming a flavor ratio of $\nu_{\mu}:\nu_{e}:\nu_{\tau}=1:1:1$ at Earth.  As discussed in Section \ref{astronu}, evidence for a diffuse astrophysical $\nu_{\mu}$ flux would manifest in the IceCube detector as a hardening at the high energy tail of the reconstructed energy observable distribution above the expectation from the atmospheric $\nu_{\mu}$ spectrum.  The energy-correlated observable used in the analysis is  the muon energy loss per unit length and is described in the next section.  

\subsection{Energy Reconstruction}

It was natural in this analysis to use the average muon energy loss per meter ($\mathrm{d}E_{\mathrm{reco}}/\mathrm{d}X$) as the energy-correlated observable since IceCube measures the muon energy loss (and not the muon energy directly) in the form of the Cherenkov photons emitted by the various stochastic muon energy loss mechanisms.  In order to estimate $\mathrm{d}E_{\mathrm{reco}}/\mathrm{d}X$ from the observed collection of Cherenkov photoelectrons  (denoted by $\{n\}$) and the expected Cherenkov photoelectron profile (denoted by $\mu$ and is explicitly a function of $\mathrm{d}E/\mathrm{d}X$), a log-likelihood based reconstruction method is used.  There is an observed $\{n\}$ and expected $\mu$ for every DOM in the detector. With an observed photoelectron collection $\{n\}$ given an expected photoelectron distribution $\mu(\mathrm{d}E/\mathrm{d}X)$ binned into $N$ bins of photoelectrons in a single DOM,  a poisson likelihood function yields:
\begin{equation}
\log L\left(\frac{\mathrm{d}E_{\mathrm{reco}}}{\mathrm{d}X}|\{n\}\right)=\sum_{i=1}^{N}n_{i}\log\mu_{i} - \mu_{i} 
\label{llh}
\end{equation}
where $n_{i}$ and $\mu_{i}$ are the observed and expected number of photoelectrons in the $i$th bin, respectively.  The Cherenkov photoelectrons are binned according to their respective arrival times at the DOM.  To obtain the total likelihood function for the detector, the log-likelihood values of the individual DOMs were summed together: 
\begin{equation}
\log L_{\mathrm{total}}=\sum_{j=1}^{\mathrm{N_{DOMs}}} \log L_{j}.
\label{llhsum}
\end{equation}

\begin{figure*}[htp]
\includegraphics[width=0.495\textwidth]{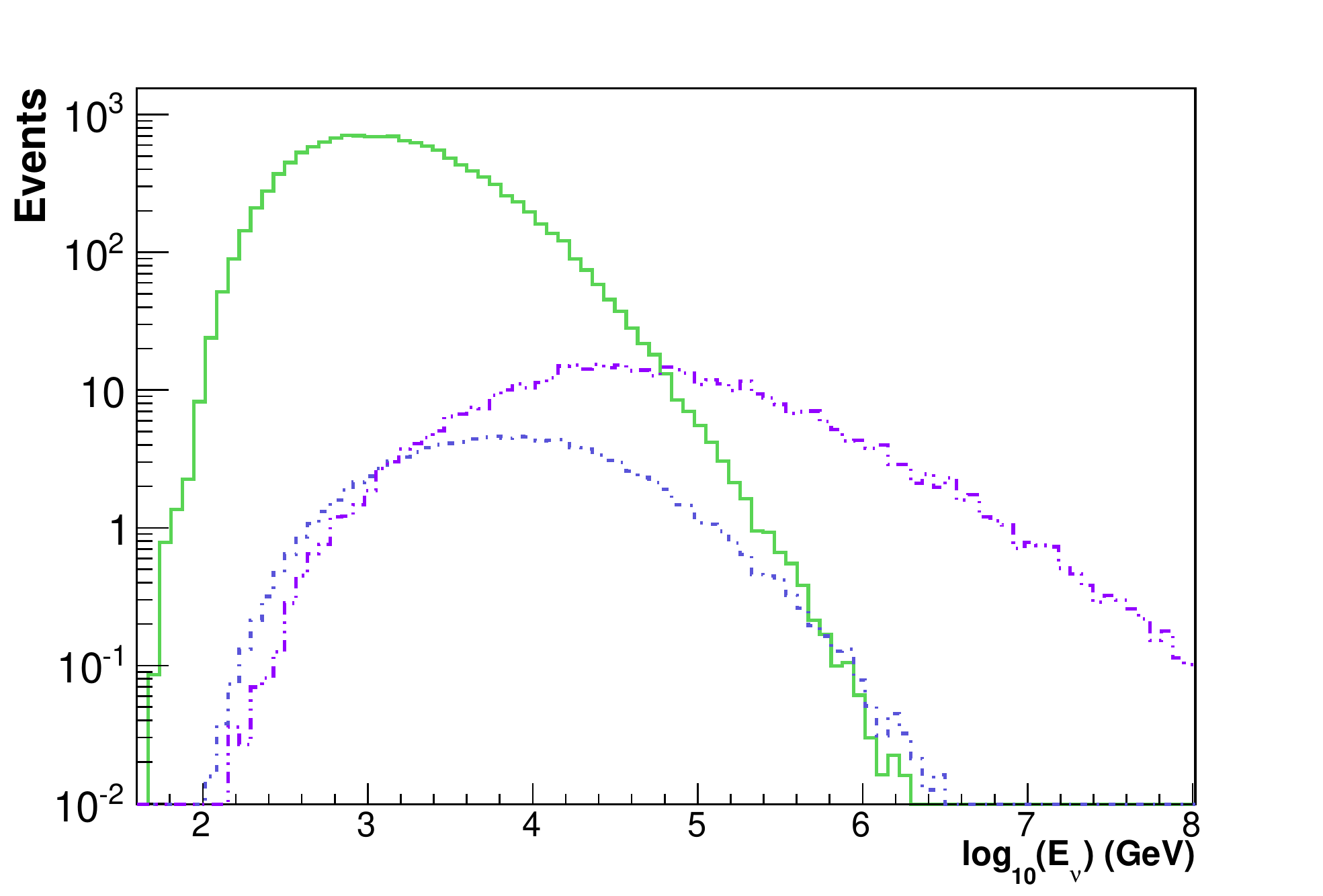}
\includegraphics[width=0.495\textwidth]{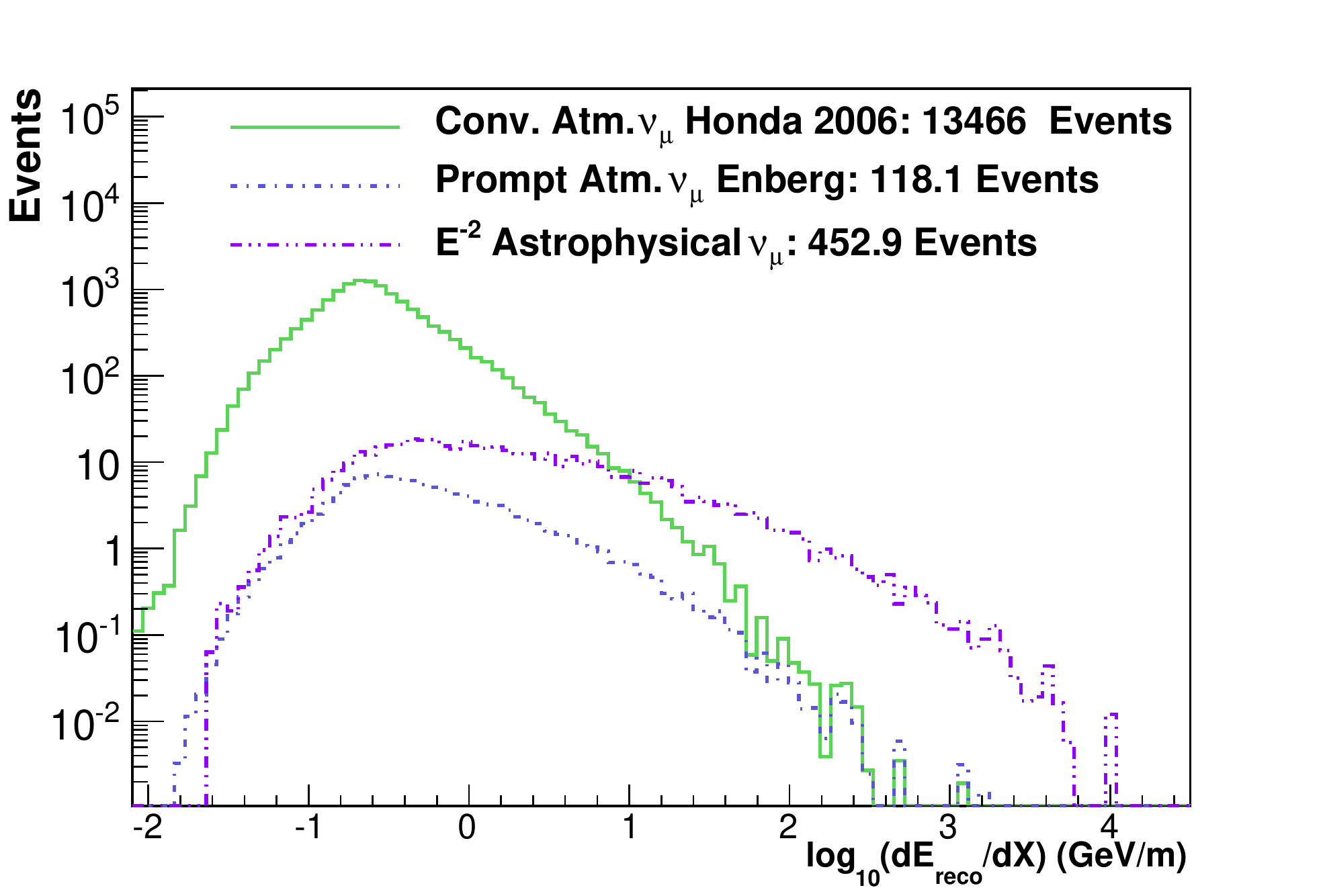}
\caption[Simulated Atmospheric and Astrophysical $\nu_{\mu}+\bar{\nu}_{\mu}$ $\mathrm{d}E_{\mathrm{reco}}/\mathrm{d}X$ Distribution]{\label{1yrmcdedx} Simulated neutrino energy distribution (left plot) and the simulated reconstructed muon energy loss distribution (right plot) of the final event sample for the Honda \textit{et. al} conventional atmospheric $\nu_{\mu}$ (green) flux model,  the Enberg \textit{et al.}  prompt atmospheric $\nu_{\mu}$ (light blue) flux model, and an astrophysical $E^{-2}$ (purple) flux with a normalization of $N_{a}=10^{-7} \mathrm{GeV \ cm^{-2} \ s^{-1} \ sr^{-1}}$.}  
\end{figure*}

The direction and geometry of the muon were fixed to the results of the MPE reconstruction.  The muon light profile $\mu(\mathrm{d}E/\mathrm{d}X)$ was parametrized in terms of the stochastic cascade energy which is varied until the likelihood function was maximized. The estimation of $\mathrm{d}E_{\mathrm{reco}}/\mathrm{d}X$ is contingent on modeling $\mu$, which depends both on the light yield of the muon and the optical properties of the South Pole ice.  Incorporating the muon light yield to the likelihood fit proved a challenge due to the stochastic energy loss processes of pair production, photo-nuclear interactions, and bremsstrahlung radiation which dominate over continuous energy losses above $\gtrsim 1$ TeV.  The relationship is approximately linear in the stochastic energy regime, however, with $\mathrm{d}E/\mathrm{d}X=a+bE$ with $a=0.25958 \ \mathrm{GeV/mwe}$ and $b=3.5709\times10^{-4} \ \mathrm{mwe^{-1}}$ in ice \cite{Rhodemuon}.  The continuous and stochastic energy losses were parametrized by the coefficients $a$ and $b$ respectively and both are written in terms of meters of water equivalent (mwe). The reconstruction algorithm therefore modeled the stochastic energy loss of a muon as uniform along the track.  

The uniform energy loss model allows one to differentiate Eq. \ref{llh} with respect to a muon energy scale factor. This leads to an analytic solution for $\mathrm{d}E_{reco}/\mathrm{d}X$ in terms of the ratio of the total observed charge across all DOMs to the total predicted charge. The $\mathrm{d}E/\mathrm{d}X$ reconstruction algorithm incorporated the optical properties of the South Pole ice into the reconstruction.  The $\mathrm{d}E/\mathrm{d}X$ reconstruction algorithm did not account for Cherenkov light from the hadronic shower initiated by the charged-current neutrino interaction and only reconstructed the energy loss due to the muon track itself.  This was not a limitation here, since the majority of events in our final sample ($\sim73\%$) are through-going tracks where the neutrino interaction occurred outside the instrumented volume of the detector. 

The performance of the $\mathrm{d}E/\mathrm{d}X$ reconstruction was characterized using simulated high energy muons and neutrinos that satisfy the analysis level selection criteria applied to the data.  Since IceCube measures the energy loss of the muon in the form of Cherenkov light from stochastic showers, we first characterized the intrinsic resolution of the  $\mathrm{d}E/\mathrm{d}X$ reconstruction by using a sample of simulated high energy muons. The correlation of $\mathrm{d}E_{\mathrm{reco}}/\mathrm{d}X$ with the muon energy closest to the center of the IceCube array is given in the left hand plot of Fig. \ref{dedxscatter}.  The relationship is linear over a large energy range. The correlation changes for energies below 1~TeV since the energy loss is no longer stochastic and the Cherenkov light output is nearly independent of energy.  The spread in $\mathrm{d}E_{\mathrm{reco}}/\mathrm{d}X$ does not vary strongly as a function of the muon energy.  The muon energy resolution over all energy decades in the stochastic energy regime above a TeV of the  $\mathrm{d}E/\mathrm{d}X$ reconstruction is $0.27$ in $\log(E)$.   The energy resolution was estimated from the $1 \sigma$ width of a Gaussian fit to the reconstructed $\mathrm{d}E/\mathrm{d}X$ distribution over all energies. The right hand plot of Fig. \ref{dedxscatter} shows a profile of the simulated neutrino energy for conventional atmospheric neutrinos and astrophysical neutrinos vs. $\mathrm{d}E_{\mathrm{reco}}/\mathrm{d}X$.  Muons of a given energy would result in a measured  $\mathrm{d}E_{\mathrm{reco}}/\mathrm{d}X$ distribution with a mean and RMS spread as indicated by the left hand plot of Fig. \ref{dedxscatter} independent of the primary energy spectrum. An estimate of the parent neutrino energy from a given measured $\mathrm{d}E_{\mathrm{reco}}/\mathrm{d}X$, however, depends on the assumed primary energy spectrum. 


The simulated $\mathrm{d}E_{\mathrm{reco}}/\mathrm{d}X$ response of the IceCube 40-string detector at final analysis level to the Honda \textit{et al.} conventional atmospheric neutrino flux \cite{honda:2006}, the Enberg \textit{et al.} prompt atmospheric neutrino flux \cite{sarcevicstd}, and a hypothetical astrophysical $E^{-2}$ flux is shown in Fig. \ref{1yrmcdedx}.  It clearly shows how the different parent spectra map into distinguishable reconstructed muon energy spectra.


\section{Analysis Method \label{analysis}}

To test the compatibility of the observed $\mathrm{d}E_{\mathrm{reco}}/\mathrm{d}X$ distribution in the data set with the hypotheses of muons arising from conventional atmospheric $\nu_{\mu}$, prompt atmospheric $\nu_{\mu}$, and astrophysical $\nu_{\mu}$ while incorporating various sources of systematic uncertainty, we incorporated the frequentist approach suggested by Feldman \cite{feldmanextension}.   The \textit{profile likelihood construction} procedure extends the original frequentist method of Feldman and Cousins \cite{feldmancousins} in order to incorporate sources of systematic uncertainties parametrized as nuisance parameters in the analysis.   

\subsection{Profile Likelihood Construction \label{profcon}}
The profile likelihood construction method results in fully frequentist confidence intervals for the physics parameters of interest (denoted by $\theta_{\mathrm{r}}$) while using values of the nuisance parameters (denoted by $\theta_{\mathrm{s}}$) that fit the data the best.  Specifically, we first constructed a Poisson likelihood function and binned our $\mathrm{d}E_{\mathrm{reco}}/\mathrm{d}X$ observable distribution into $N$ bins: \begin{equation}
L(\{n_{i}\}|\{\mu_{i}(\theta_{\mathrm{r}},\theta_{\mathrm{s}})\})=\prod_{i=1}^N\frac{e^{-\mu_{i}}}{n_{i}!} \mu_{i}^{n_{i}}
\label{poissllh2}
\end{equation}  where $n_{i}$ and $\mu_{i}$ denote the observed and expected event counts in the $i$th $\mathrm{d}E_{\mathrm{reco}}/\mathrm{d}X$ bin respectively.  We then iterated over the physics parameter space and calculated the profile likelihood \cite{statsbook} ratio test statistic $R_{\mathrm{p}}$ at each point $\theta_{\mathrm{r}}$.  Defining $\mathcal{L}=-2\log(L)$:  \begin{align}
R_{\mathrm{p}}(\theta_{\mathrm{r}}) = & \ \mathcal{L}(\theta_{\mathrm{r}},\hat{\hat{ \theta}}_{s}) - \mathcal{L}(\hat{\theta}_{\mathrm{r}},\hat{\theta}_{\mathrm{s}}) \\
\label{llhratiosys}
= & -2 \log \left(\frac{L(\{n_{i}\} | \{\mu_{i}(\theta_{\mathrm{r}},\hat{\hat{ \theta}}_{s}) \}) } {L(\{n_{i}\} |\{\mu_{i}(\hat{\theta}_{\mathrm{r}},\hat{\theta}_{\mathrm{s}}) \} ) }  \right) \nonumber \\
= & \  2 \sum_{i=1}^{N}\left( \mu_{i}  - \hat{\mu}_{i}+  n_{i}\log \frac{\mu_{i}}{\hat{\mu}_{i}}\right)  \nonumber
\end{align} where $\hat{\mu}_{i}=\mu_{i}(\hat{\theta}_{\mathrm{r}},\hat{\theta}_{\mathrm{s}})$ and $\mu_{i}=\mu_{i}(\theta_{\mathrm{r}},\hat{\hat{ \theta}}_{s})$. $\hat{\theta}_{\mathrm{r}}$ and $\hat{\theta}_{\mathrm{s}}$ denote the values of the physics and nuisance parameters that globally minimize the profile likelihood $\mathcal{L}(\hat{\theta}_{\mathrm{r}},\hat{\theta}_{\mathrm{s}})$ and therefore describe the data the best. The value of the nuisance parameter $\hat{\hat{\theta}}_{\mathrm{s}}$ conditionally minimizes the profile likelihood $\mathcal{L}(\theta_{\mathrm{r}},\hat{\hat{\theta}}_{\mathrm{s}})$ at the physics parameter point $\theta_{\mathrm{r}}$.  The profile likelihood test statistic was marginalized over nuisance parameters in the likelihood ratio such that confidence intervals were constructed solely for the physics parameters of interest.

Confidence intervals were constructed at confidence level $\alpha$ by comparing the profile likelihood test statistic to a critical value $R_{\mathrm{p,crit}}$ at each point $\theta_{\mathrm{r}}$. The critical profile likelihood value determines if a hypothesis is accepted or rejected at a certain confidence level.  Following the prescription outlined in Refs. \cite{feldmanextension} and \cite{feldmancousins}, we defined $R_{\mathrm{p,crit}}(\theta_{\mathrm{r}})$ by examining the spread of the profile likelihood test statistic $R_{\mathrm{p}}(\theta_{\mathrm{r}})$ caused by statistical fluctuations.  This was facilitated by generating a number of Monte Carlo experiments to obtain the distribution of $R_{\mathrm{p}}(\theta_{\mathrm{r}})$ at each physics point $\theta_{\mathrm{r}}$ while fixing the values of the nuisance parameters to $\hat{\hat{\theta}}_{\mathrm{s}}$.  Confidence intervals were constructed at confidence level $\alpha$ by finding the critical value of the profile likelihood, $R_{\mathrm{p,crit}}(\theta_{\mathrm{r}})$, such that the fraction $\alpha$ of experiments at $\theta_{\mathrm{r}}$ satisfied $R_{\mathrm{p}}(\theta_{\mathrm{r}}) < R_{\mathrm{p,crit}}(\theta_{\mathrm{r}})$.  The acceptance region is the parameter space $\{\theta_{\mathrm{r}}\}$ such that $R_{p,data}(\theta_{\mathrm{r}}) < R_{\mathrm{p,crit}}(\theta_{\mathrm{r}})$ at a chosen confidence level $\alpha$.  By utilizing the profile likelihood distribution to determine the confidence level, we have used the likelihood ratio as an ordering principle in order to sort the possible experimental outcomes by increasing statistical significance.    

\subsection{Systematic Uncertainties}

\begin{table*}[htp]
\caption{\label{systematicerrorlist1}Systematic uncertainties in the atmospheric neutrino flux and the detector response that affect the shape and overall normalization of the $\mathrm{d}E_{\mathrm{reco}}/\mathrm{d}X$ observable.  The effect on the rate of triggered atmospheric $\nu_{\mu}$ is shown in the third column.}
\begin{tabular}{l l l}
\hline \hline
Systematic Uncertainty & Magnitude & Atm. $\nu_{\mu}$ Rate \\ [0.5ex]
\hline
Conv. Atmospheric $\nu_{\mu}+\bar{\nu}_{\mu}$ Flux & $\pm25\%$ & $\pm25\%$ \\
Prompt Atmospheric $\nu_{\mu}+\bar{\nu}_{\mu}$ Flux & $-44\%,+25\%$ &  $-44\%,+25\%$ \footnote{The asymmetric error range in the overall flux normalization of the prompt atmospheric neutrino flux only affects the overall rate of prompt atmospheric neutrinos} \\
Cosmic Ray Spectral Slope & $\pm0.03 $ & negligible\\
DOM Sensitivity & $\pm 8 \%$ & $\pm 15\%$ \\
Scattering \&Absorption Coefficients &  $\pm 10\%$ & $-13.5\%,+14.2\%$ \\
Neutrino-Nucleon Cross Section &  $\pm 3\%$ & $\pm 3\%$ \\
Muon Energy Loss &  $\pm 1\%$ & negligible \\
Bedrock Density &  $\pm 10\%$ & negligible  \\
\hline
\end{tabular}
\label{systematicerrorlist1}
\end{table*}
\footnotetext{The asymmetric error range in the overall flux normalization of the prompt atmospheric neutrino flux only affects the overall rate of prompt atmospheric neutrinos}
Systematic errors were incorporated into the analysis as nuisance parameters in the likelihood. The profile likelihood construction method discussed in the last section was used to define confidence regions for the physics parameters of interest while incorporating the various sources of systematic uncertainty outlined in this section.  This work considered sources of systematic errors that affect the shape and overall normalization of the observed $\mathrm{d}E_{\mathrm{reco}}/\mathrm{d}X$ distribution. The sources of systematic uncertainty are summarized in Table \ref{systematicerrorlist1}.

One of the largest sources of systematic uncertainty is the overall normalization of the atmospheric neutrino flux.  The model used in this analysis for the conventional component of the atmospheric neutrino flux was derived by Honda \textit{et al.} \cite{honda:2006}, where the uncertainty in the absolute normalization was estimated to be $\pm25\%$.  The prompt component of the atmospheric neutrino flux has yet to be experimentally measured and there exists a large range in the overall normalization as calculated in various model predictions.  The baseline model used in this analysis for the prompt component of the atmospheric neutrino flux is the calculation from Enberg \textit{et al.} \cite{sarcevicstd} where the authors quoted a standard perturbative QCD model prediction with an asymmetric error range in the overall flux normalization of $-44\%$ to $+25\%$.  

The spectral shape of the atmospheric neutrino flux is affected by the uncertainty in the spectral slope of the primary cosmic ray spectrum. The uncertainty in the primary cosmic ray spectrum was estimated by considering the uncertainty in the spectral slopes of cosmic ray protons (which comprise $79\%$ of the flux) and of helium nuclei ($15\%$ of the flux).  The remaining $6\%$ predominately consists of elements heavier than helium. Gaisser \textit{et al.} \cite{GaisserSlope} estimated the spectral slope uncertainty for protons to be $\pm0.01$ and for helium nuclei to be $\pm0.07$.  Scaling the individual spectral index uncertainties by the fraction of the total flux for the respective component gave an uncertainty in the primary cosmic ray spectral slope of $\pm0.03$. This range is valid for a primary cosmic ray energy up to about $1$ PeV. Above this energy appears a region between $1$ and $10$ PeV  known as the knee where the primary cosmic ray spectrum steepens \cite{GaisserCosmicRays}.  The cosmic ray composition around the knee is still an active area of research.  The nominal prediction of the model calculated in Ref. \cite{honda:2006} was calculated up to a neutrino energy of only $10$ TeV.  We extrapolated the Honda \textit{et al.} model beyond this energy by assuming the model followed its approximate $E^{-3.7}$ shape above $10$ TeV.  We note that such an extrapolation does not reflect the steepening in the spectrum expected above 1 PeV/nucleon as a consequence of the knee. 

There are two main sources of systematic uncertainties which affect the response of the IceCube detector to Cherenkov light.  The first is the uncertainty in the absolute sensitivity of the digital optical module. The DOM sensitivity was assumed to be dominated by the uncertainty in the absolute efficiency of the photomultiplier tube which was measured to be $\pm 8\%$ \cite{pmtpaper}.  The DOM sensitivity is further reduced by a shadowing effect from the main cable and the magnetic shield in the DOM, which reduces the sensitivity by  $7\%$.  The second dominant source of systematic uncertainty affecting the detector response is the uncertainty in the measured properties of the glacial ice at the South Pole.  The measured uncertainty in the scattering and absorption coefficients of the south pole ice was measured to be $\pm 10\%$ \cite{spicepaper} at a flasher LED wavelength of $405$ nm. 

Other relatively minor sources of systematic error were quantified with dedicated simulation studies.  The uncertainty in the charged current, deep-inelastic neutrino-nucleon cross section was calculated in  \cite{amandapointsource} to be $\pm3\%$ using the parton distribution function error tables from \cite{cteq5} and the error calculation prescription in Ref. \cite{cteqerror}.  The 3$\%$ uncertainty in the cross section corresponds to a $3\%$ uncertainty in the overall neutrino event rate.  The uncertainty in the muon energy loss cross sections was estimated from \cite{mmc} to be $1\%$ which has a negligible effect on the overall event rate since a decrease or increase in the stochastic cross sections are accompanied by a corresponding increase or decrease in the muon range. The uncertainty in the density of the bedrock under the polar ice was measured to be $10\%$ \cite{rockerror}.  This provided a negligible difference in the atmospheric neutrino event rates of $ < 0.1 \%$, since the increase in the neutrino interaction probability is offset by a corresponding decrease in the range of the muon.  The background contamination in the final event sample was estimated to be less than $1\%$, and was therefore a negligible source of systematic uncertainty in the analysis.

\subsection{Final Analysis Parameters}

The systematic errors summarized in Table \ref{systematicerrorlist1} were incorporated into the profile likelihood as nuisance parameters.  During minimization, each nuisance parameter was allowed to vary freely within the allowed range around its nominal value.  The nominal values of the nuisance parameters correspond to the predictions of the Honda \textit{et al.} model for the conventional atmospheric flux and baseline values as given by simulation for the other nuisance parameters. Each point in the likelihood space gave a specific prediction for the $\mathrm{d}E_{\mathrm{reco}}/\mathrm{d}X$ observable and the profile construction method was used to define confidence regions for the physics parameters of interest.  

The likelihood methodology could be used for two main analyses. The primary analysis is the search for a diffuse astrophysical $\nu_{\mu}$ signal while simultaneously fitting for a potential prompt component of the atmospheric $\nu_{\mu}$ flux.  In the absence of any signal, the profile likelihood construction could be used to measure the conventional atmospheric neutrino flux. 

The generic astrophysical diffuse $\nu_{\mu}$ flux was parametrized as an $E^{-2}$ spectrum and the normalization is the main physics parameter of interest: \begin{equation}
\Phi_{\mathrm{a}}=N_{\mathrm{a}}E^{-2}
\end{equation} where $N_{\mathrm{a}}$ has units of $\mathrm{GeV \ cm^{-2} \ s^{-1} \ sr^{-1}}$.  Astrophysical models that do not predict an $E^{-2}$ spectral shape were also considered in this work.  The second physics parameter of interest denotes the absolute normalization of the prompt atmospheric neutrino flux:  \begin{equation}
\Phi_{\mathrm{p}}=(1+\alpha_{\mathrm{p}})\left(\frac{E}{E_{\mathrm{median,p}}}\right)^{\Delta\gamma}\Phi_{\mathrm{Enberg}}
\label{promptphi}
\end{equation}  where $1+\alpha_{\mathrm{p}}$ describes the deviation in the absolute normalization of the prompt atmospheric neutrino flux from the reference prompt atmospheric neutrino model which was taken to be the calculation from Enberg \textit{et al}.   The uncertainty in the primary cosmic ray slope, $\Delta\gamma$, changes the shape of the predicted atmospheric neutrino flux and was a nuisance parameter in the analysis.  It was allowed to float in the $\pm 0.03$ range quoted in Table \ref{systematicerrorlist1}. The shape dependent term was modeled by introducing an energy dependent weight $(E/E_{\mathrm{median,p}})^{\Delta\gamma}$ where $E_{\mathrm{median,p}}$ is the median neutrino energy at final purity level.  The median energy is $7$ TeV for the nominal prediction by Enberg \textit{et al.}  


The conventional component of the atmospheric neutrino flux was treated as a nuisance parameter in the main analysis and was parametrized in the same fashion as the prompt flux: \begin{equation}
\Phi_{\mathrm{c}}=(1+\alpha_{\mathrm{c}})\left(\frac{E}{E_{\mathrm{median,c}}}\right)^{\Delta\gamma} \Phi_{\mathrm{Honda}}
\label{convphi}
\end{equation}  where $1+\alpha_{\mathrm{c}}$ describes the deviation in the absolute normalization of the conventional atmospheric neutrino flux from the reference model by Honda \textit{et al.},   $\Delta\gamma$ is again the uncertainty in the primary cosmic ray slope, and  $E_{\mathrm{median,c}}$ is  the median conventional atmospheric neutrino energy at final purity level.  A shape dependent term was again modeled by an energy dependent weight $(E/E_{\mathrm{median,c}})^{\Delta\gamma}$ where $E_{\mathrm{median,c}}$ is $1.2$ TeV for the nominal prediction by Honda \textit{et al.} 

The detector efficiency, denoted by $\epsilon$, affects the overall event rate in the IceCube detector.  The magnitude of this systematic error combines in quadrature the systematic uncertainties in the absolute DOM sensitivity, the neutrino interaction cross section, the muon energy loss cross sections, and the bedrock density giving an allowed range of $\pm 8.3\%$.  The detector efficiency nuisance parameter was implemented by assuming that the absolute DOM sensitivity is independent of energy.   Although the uncertainty in the absolute DOM sensitivity affects the event rate for lower energy neutrino events more than higher energy events, this energy dependence was neglected since the primary astrophysical diffuse search is dominated by the high energy tail of the $\mathrm{d}E_{\mathrm{reco}}/\mathrm{d}X$ distribution. 


The scattering and absorption coefficients $b(\lambda=405\mathrm{nm})$ and $a(\lambda=405\mathrm{nm})$ (both measured at a LED wavelength of $405 \mathrm{nm}$ \cite{spicepaper,flashersicrc}) were implemented as discrete nuisance parameters in the likelihood function.   This was facilitated by generating Monte Carlo neutrino simulation sets with the scattering and absorption coefficients increased by $10\%$, decreased by $10\%$, the scattering increased and absorption decreased by $10\%$, and the scattering decreased and absorption increased by $10\%$.

\begin{table}[hpt]
\caption{\label{analysislist}Physics parameters and nuisance parameters used for the astrophysical diffuse $\nu_{\mu}$ search and the measurement of the atmospheric neutrino spectrum}
\begin{tabular}{c c c}
\hline \hline
Analysis & Physics Parameters & Nuisance Parameters \\ [0.5ex]
\hline
Astro $\nu_{\mu}$  & $N_{\mathrm{a}}$,\ $1+\alpha_{\mathrm{p}}$ & $1+\alpha_{\mathrm{c}}$,\ $\Delta\gamma$,\ $\epsilon$,\ $b(405)$,\ $a(405)$\\
Atm.  $\nu_{\mu}$ & $1+\alpha_{\mathrm{c}}$, $\Delta\gamma$ & $\epsilon$, $b(405)$, $a(405)$\\
\hline
\end{tabular}
\end{table}

To summarize, the profile likelihood used in the main analysis incorporated two physics parameters and five nuisance parameters.  In the absence of any signal, the  conventional atmospheric neutrino flux measurement promotes the deviation in the conventional atmospheric flux  $1+\alpha_{\mathrm{c}}$ and the uncertainty in the primary cosmic ray spectral slope $\Delta\gamma$ to physics parameters, giving a likelihood with two main physics parameters and three nuisance parameters.  The physics and nuisance parameters for the two analyses are summarized in Table \ref{analysislist}.  

\subsection{Sensitivity and Discovery Potential}


A blindness procedure was followed in order to prevent any inadvertent tuning of the purity cuts that would bias the analysis.  To establish a context for the unblinded results, we quantified the limit setting potential of the analysis (the analysis sensitivity) and the ability  to discover an astrophysical neutrino flux (the discovery potential).  Specifically, the sensitivity is defined as the median $90\%$ upper limit obtained over an ensemble of simulated experiments with no true signal. The discovery potential is defined to be the strength a hypothetical astrophysical $\nu_{\mu}$ flux required to obtain a $5\sigma$ discovery in $90\%$ of simulated experiments in the ensemble.  The sensitivity of this analysis to a diffuse flux of astrophysical $\nu_{\mu}$ with an $E^{-2}$ spectrum is $1.22\times10^{-8} \ \mathrm{GeV \ cm^{-2} \ s^{-1} \ sr^{-1}}$ and the $E^{-2}$ discovery potential is $6.1\times10^{-8} \ \mathrm{GeV \ cm^{-2} \ s^{-1} \ sr^{-1}}$.

\section{Results \label{results}}

\begin{figure}[htp]
\includegraphics[width=0.5\textwidth]{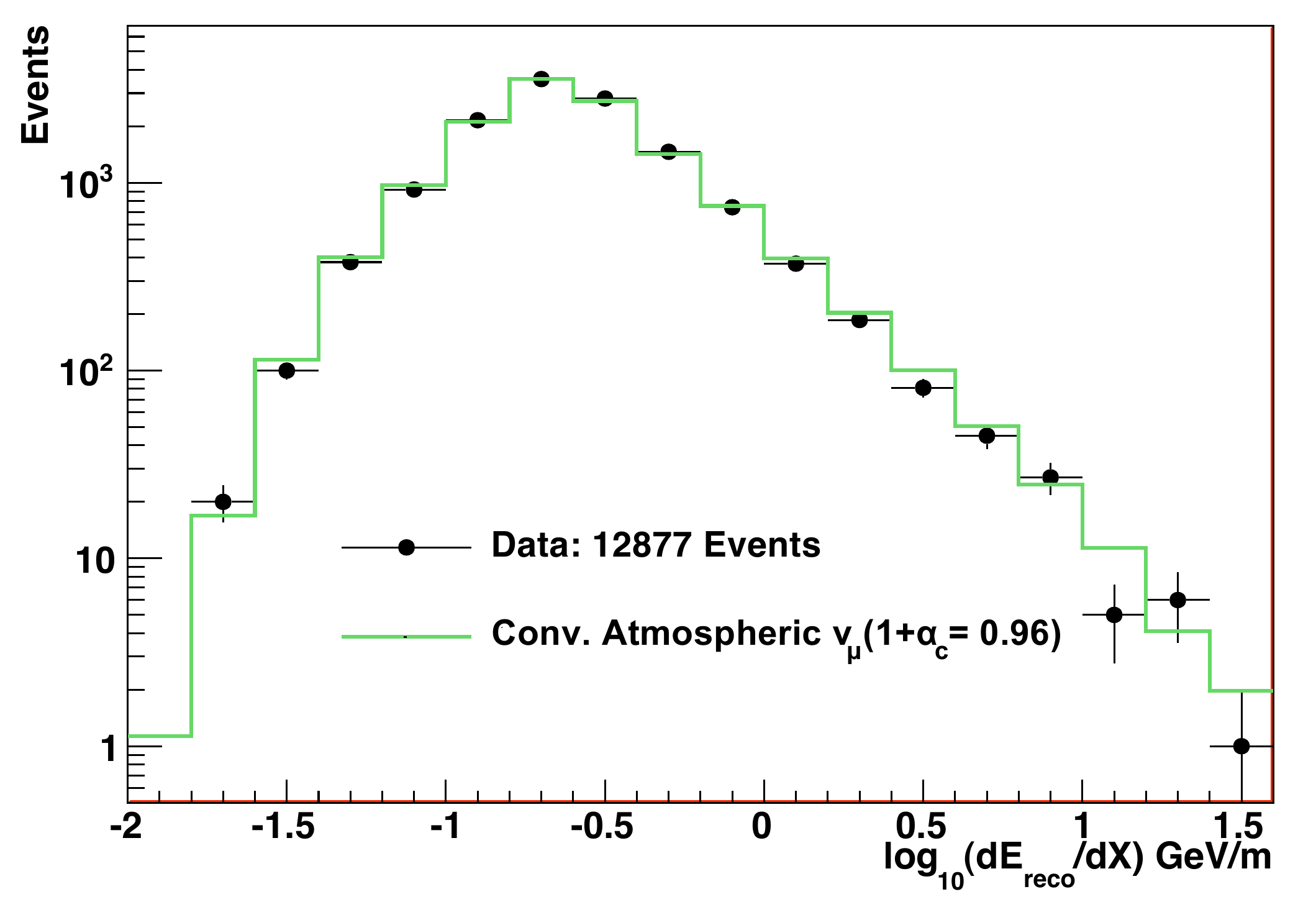}
\caption[$\mathrm{d}E_{\mathrm{reco}}/\mathrm{d}X$ Fitted to the Data]{\label{fitteddedx}The fitted muon energy loss distribution of the final event sample is shown.  The best fit to the data (black, shown with $1 \sigma$ error bars) consists only of conventional atmospheric $\nu_{\mu}$, and no evidence is found for a prompt atmospheric $\nu_{\mu}$ flux or an astrophysical $E^{-2}$ $\nu_{\mu}$ flux. }  
\end{figure}

After we performed the profile likelihood construction analysis on the $\mathrm{d}E_{\mathrm{reco}}/\mathrm{d}X$ distribution, no evidence was found for an astrophysical neutrino flux or a prompt component of the atmospheric neutrino flux.  The fitted $\mathrm{d}E_{\mathrm{reco}}/\mathrm{d}X$ distribution is shown in Fig. \ref{fitteddedx} and the best fit values of the analysis parameters to the data are summarized in Table \ref{fitresults}.  No evidence for a signal was seen,  so upper limits were set for astrophysical neutrino flux models.

\begin{table*}[htp]
\caption{\label{fitresults}Likelihood fit results and associated errors reported by the fit.  Errors are quoted as $1\sigma$ unless otherwise noted. The allowed range of the nuisance parameters are also given as $1\sigma$ Gaussian constraints.}
\centering
\begin{tabular}{c c c c}
\hline \hline
Parameter & Best Fit & Error & Constrained Range \\ [0.5ex]
\hline
$N_{\mathrm{a}}$ & $0$ & $8.9\times10^{-9} \ \mathrm{\frac{GeV}{cm^{2} \ s \  sr}}$ ($90\%$ U.L.) & N/A\\
$1+\alpha_{\mathrm{p}}$ & $0$ & $0.73$ ($90\%$ U.L.) & N/A \\
$1+\alpha_{\mathrm{c}}$ & $0.96$ & $\pm0.16$ & $\pm0.25$\\
$\Delta\gamma$ & $-0.032$ & $\pm0.014 $ & $\pm0.03$\\
$\epsilon$ & $+2\%$ &$\pm 8.3\%$ & $\pm8.3\%$ \\
$b_{\mathrm{e}}(\lambda=405\mathrm{nm})$ & Nominal & $\pm10\%$ & $\pm10\%$ \\
$a(\lambda=405\mathrm{nm})$ & Nominal & $\pm10\%$ & $\pm10\%$ \\
\hline
\end{tabular}
\end{table*}

\subsection{Upper Limits on Astrophysical Neutrino Fluxes}

\begin{figure}[hbtp]
\centering
\includegraphics[width=0.515\textwidth]{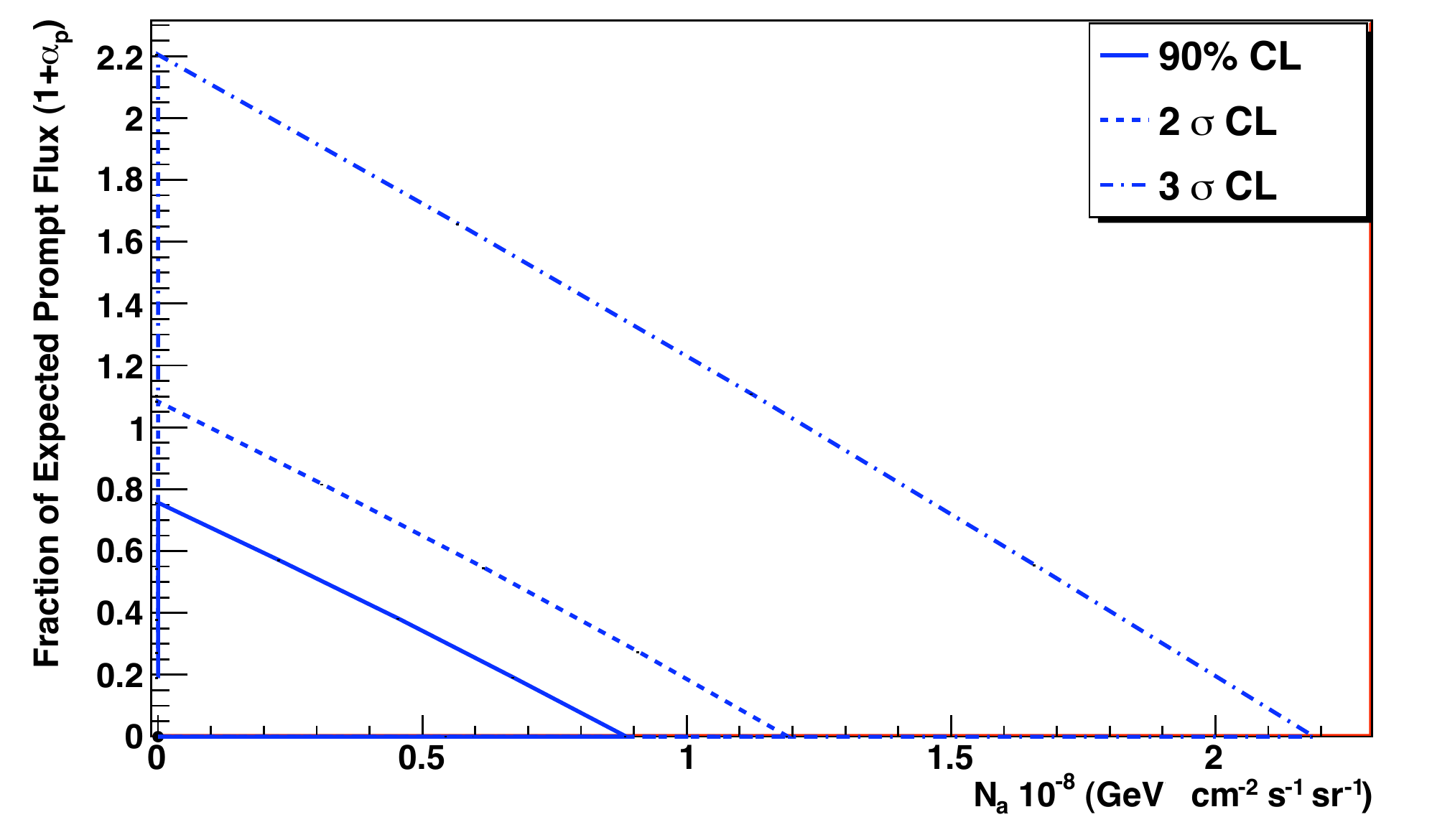}
\caption[Allowed Regions for Astrophysical $E^{-2} \ \nu_{\mu}$ and Prompt Atmospheric $\nu_{\mu}$]{\label{upperlimitfc}Allowed regions for astrophysical muon neutrinos with an $E^{-2}$ spectrum and prompt atmospheric neutrinos at $90\%$, $2\sigma$, and $3\sigma$ confidence level.  The lines indicate the boundary of the allowed region at the stated confidence level. The best fit point is indicated by the black dot at the origin.}  
\end{figure}

\begin{figure*}[htp]
\includegraphics[scale=0.93]{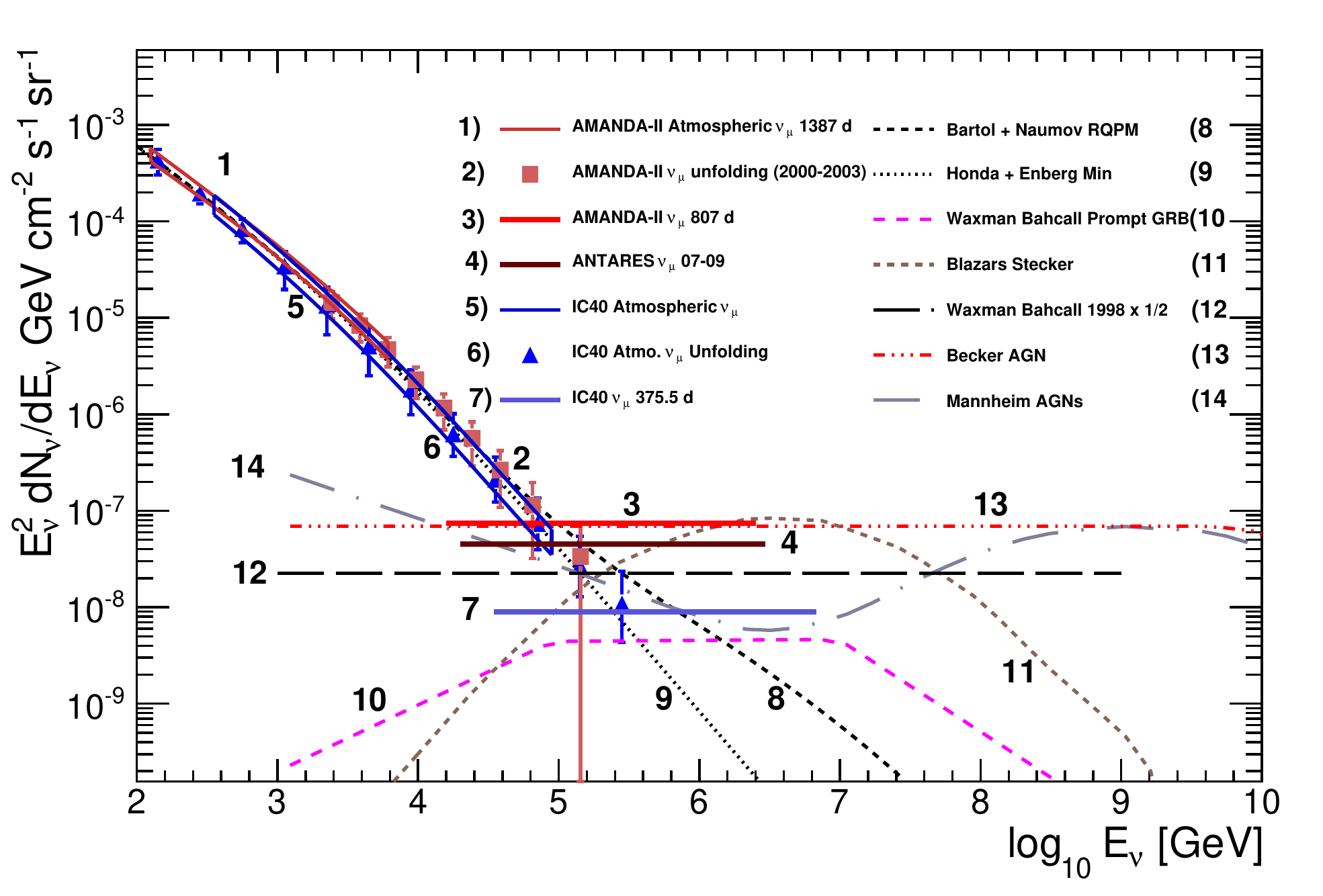}
\caption[Upper Limits on Astrophysical Neutrino $\nu_{\mu}$ Fluxes]{\label{singleflavorlimit}Upper limits on an astrophysical $\nu_{\mu}$ flux with an $E^{-2}$ spectrum are shown along with theoretical model predictions of diffuse astrophysical muon neutrinos from different sources.  The astrophysical $E^{-2} \ \nu_{\mu}$ upper limits shown are from AMANDA-II \cite{jesspaper}, ANTARES \cite{antareslimit}, and the current work utilizing the IceCube 40-string configuration (IC40 $\nu_{\mu} \ 375.5$ d).  The atmospheric $\nu_\mu$ measurements shown are from AMANDA-II \cite{kelleypaper,kierstenpaper}, the IceCube 40-string (IC40) unfolding measurement \cite{warrenpaper} and the current work (IC40 Atmospheric $\nu_{\mu}$).  }  
\end{figure*}

The allowed regions for the normalization of the astrophysical flux $N_{\mathrm{a}}$ corresponding to an $E^{-2} \ \nu_{\mu}$ flux  and the normalization for prompt atmospheric neutrinos are shown in Fig. \ref{upperlimitfc}.  The upper limit for the astrophysical normalization $N_{\mathrm{a}}$ at  $90\%$ confidence level was obtained from Fig. \ref{upperlimitfc} by finding the point on the $90\% \  C.L.$  boundary along the null hypothesis of no prompt atmospheric neutrinos.   The $90\%$ upper limit on a hypothetical astrophysical $\Phi_{\nu_{\mu}}=N_{\mathrm{a}} E^{-2}$ flux at Earth with systematic uncertainties included is $N_{\mathrm{a}}^{90\%}=8.9 \times 10^{-9} \ \mathrm{GeV \ cm^{-2} \ s^{-1} \ sr^{-1}}$.   The analysis is sensitive in the energy range between $35 \ \mathrm{TeV}$ and $7 \ \mathrm{PeV}$.  The energy range was determined from MC simulation studies of the analysis sensitivity, which was calculated to be $1.22\times10^{-8} \ \mathrm{GeV \ cm^{-2} \ s^{-1} \ sr^{-1}}$.  The energy range was determined by introducing an energy threshold and ceiling such that the analysis sensitivity changed by five percent. The $90\%$ upper limit derived in this work on a hypothetical astrophysical $\nu_{\mu}$ flux is compared to other $\nu_{\mu}$ limits and flux models in Fig. \ref{singleflavorlimit}.

\begin{table*}[htbp]
\caption{\label{astroresults}Upper Limits for Astrophysical $\nu_{\mu}$ for different Astrophysical Models. The upper limits are expressed in terms of the model rejection factor \cite{mrf}, which is  the percentage of the reference model rejected at the stated confidence level such that $\Phi_{\mathrm{C.L}}=\mathrm{MRF}\times\Phi_{\mathrm{ref}}$.}
\centering
\begin{tabular}{c c c c c c}
\hline \hline
Model & $90\%$ C.L. & $3\sigma$ C.L. & $5\sigma$ C.L & 90$\%$ Energy Range (TeV-PeV)\\ [0.5ex]
\hline
$E^{-2} \ \left(\mathrm{\frac{GeV}{cm^{2}\ s \  sr}}\right)$ & $0.89 \times 10^{-8}$ & $2.2 \times 10^{-8}$& $4.0 \times 10^{-8}$ &  $35  - 7 $   \\
W-B  Upper Bound & $0.4$ & $0.97$&  $1.78$ & $35 - 7 $ \\
Stecker Blazar & $0.1$  & $0.32$ & $0.42$ & $120 - 15 $  \\
BBR FSRQ & $0.12$  & $0.34$ & $0.46$ & $35  - 7 $\\
Mannheim AGN & $0.05$ & $0.21$ & $0.4$ & $9  - 1 $  \\
\hline
\end{tabular}
\end{table*}

\begin{figure*}[htp]
\includegraphics[scale=0.95]{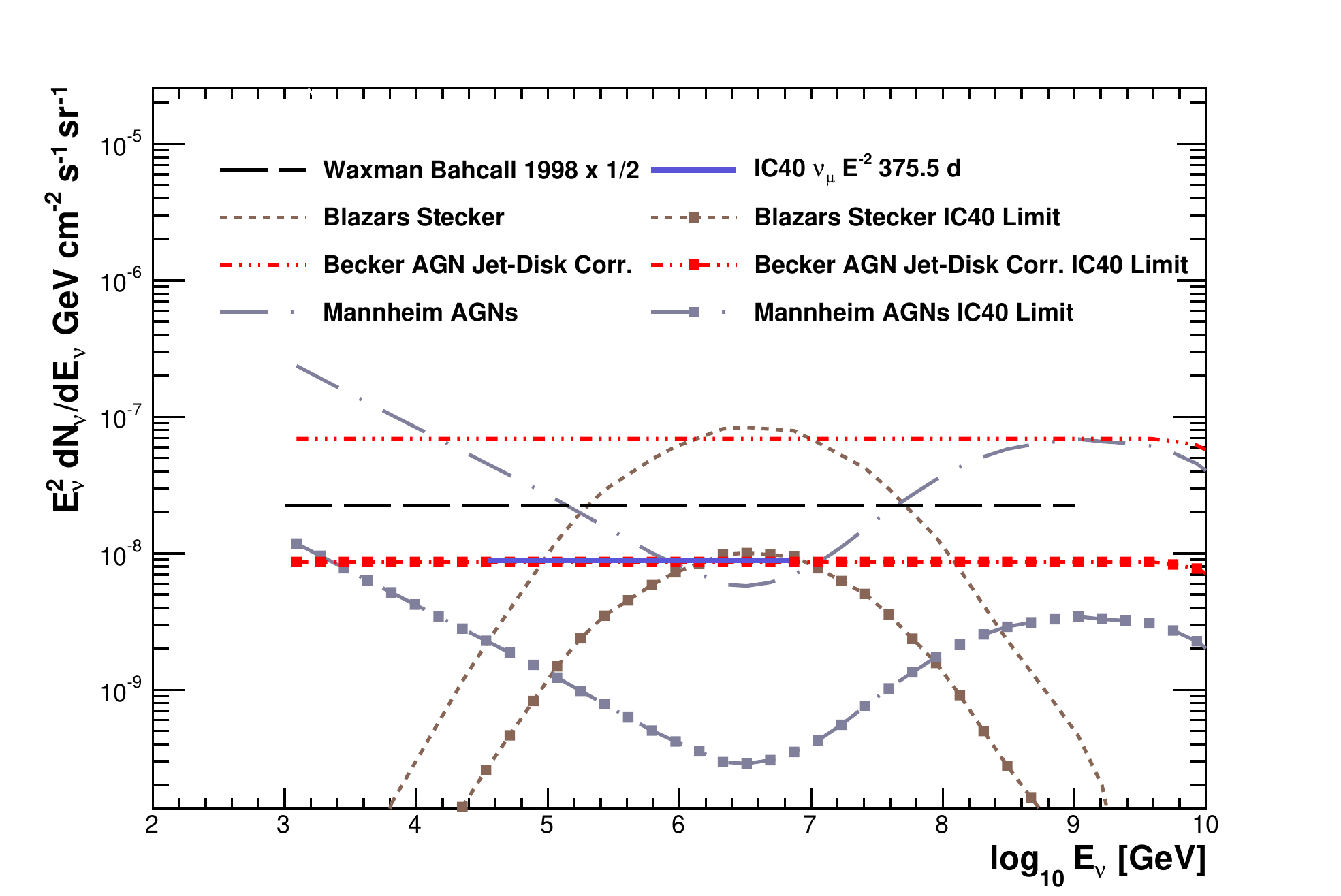}
\caption[Upper Limits on Astrophysical Neutrino $\nu_{\mu}$ Flux Models]{\label{fluxeslimit}Upper limits derived by this work utilizing the IceCube 40-string configuration (IC40) are shown for an $E^{-2}$ astrophysical $\nu_{\mu}$ flux and theoretical model predictions that do not predict an $E^{-2}$ spectrum. }  
\end{figure*}

Astrophysical neutrino models that do not predict an $E^{-2}$ spectrum from various source classes were tested in the analysis.  Of the models considered, this analysis was sensitive to the blazar model derived by Stecker \cite{steckerblazar}, the AGN neutrino model derived by Mannheim \cite{mannheimagn}, and the radio galaxy neutrino model from Becker, Biermann, and Rhode \cite{beckerfsrq}.  These models were rejected at the $5\sigma$ confidence level.  The analysis also rules out the Waxmann-Bahcall upper bound \cite{wbbound} at a $3\sigma$ confidence level.  The upper limits on astrophysical $\nu_{\mu}$ for the different models are summarized in table \ref{astroresults}.  The upper limits for the models are expressed in terms of the model rejection factor \cite{mrf}, which in the context of this analysis is  the percentage of the reference model rejected at the stated confidence level, such that $\Phi_{\mathrm{C.L}}=\mathrm{MRF}\times\Phi_{\mathrm{ref}}$.  The $90\%$ upper limits on these flux models are shown in Fig. \ref{fluxeslimit}. 
We note that the radio galaxy neutrino model from Becker, Biermann, and Rhode rejected at a $5\sigma$ confidence level was derived with a primary proton cosmic ray energy spectrum of $E^{-2}$ and an optical thickness $\tau=0.2$.  The authors summarized in Ref. \cite{beckerfsrq} a range of neutrino flux models with different spectral shapes of the primary proton cosmic ray spectrum and varying optical thickness which are below the sensitivity of these results.

\subsection{Upper Limits on Prompt Atmospheric Neutrino Flux Models}

\begin{figure*}[htp]
\includegraphics[width=0.49\textwidth]{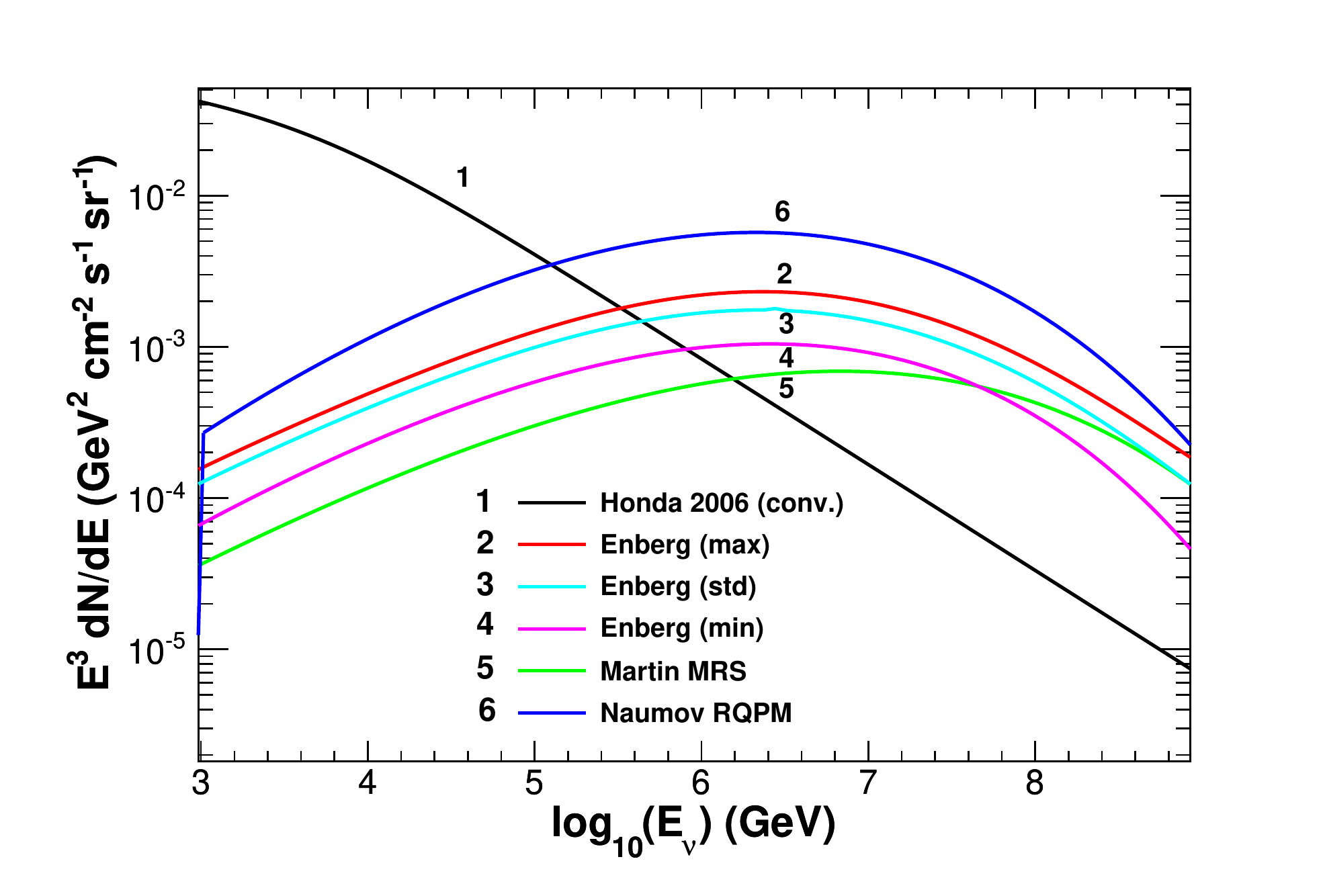}
\includegraphics[width=0.49\textwidth]{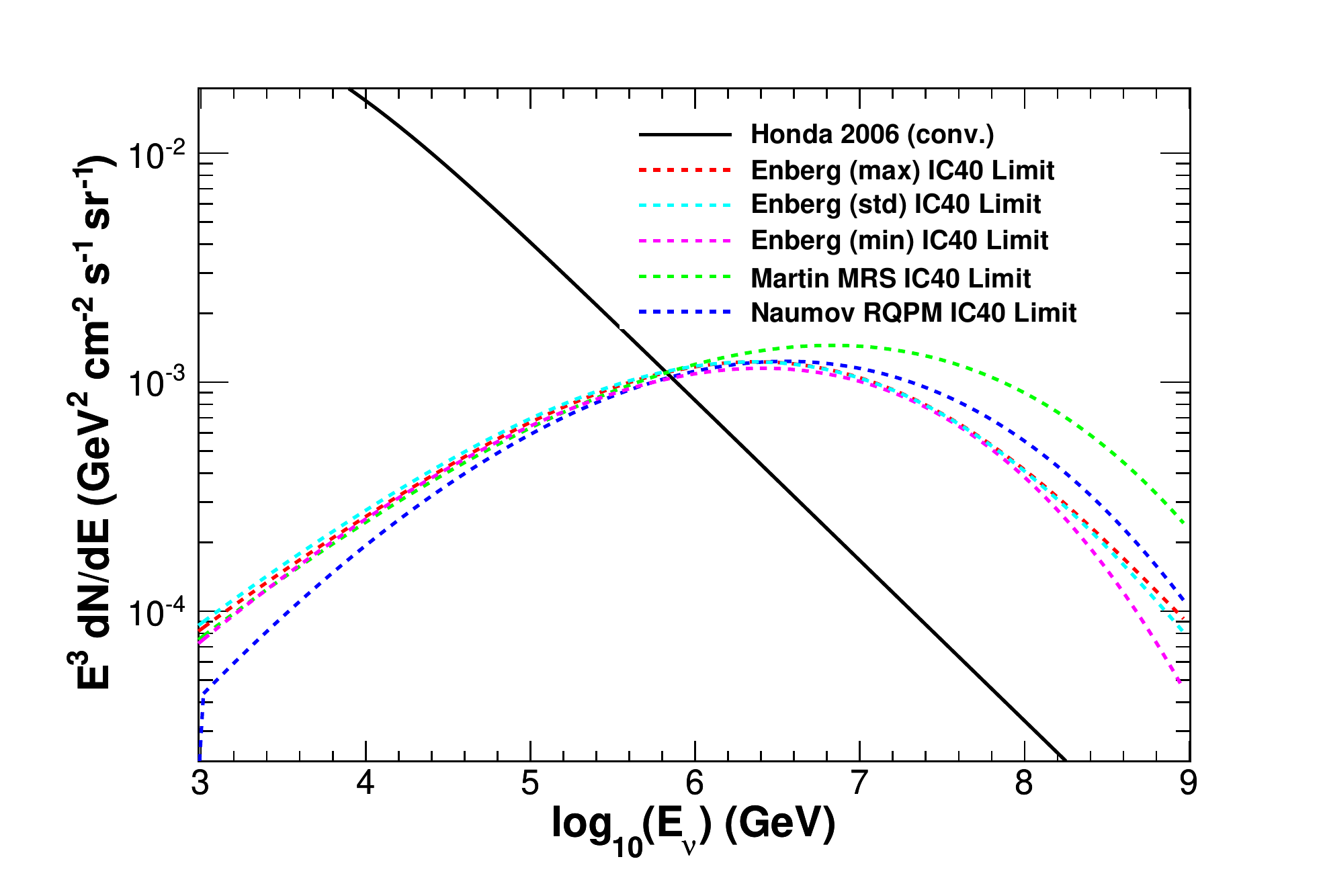}
\caption[Models of the Prompt Atmospheric Neutrino Energy Spectrum]{The left plot shows the predicted prompt atmospheric neutrino fluxes averaged over zenith angle and multiplied by $E^{3}$ to enhance features. The Honda 2006 model is shown for comparison. The right plot shows the $90\%$ confidence level upper limit on the prompt models.}  
\label{promptatmofluxes}
\end{figure*}

This analysis showed no evidence for a prompt component to the atmospheric neutrino flux.  Hypotheses other than the reference model from Enberg \textit{et al.} are shown in the left hand side in  Fig. \ref{promptatmofluxes} and were tested in this analysis.  The results of the prompt model tests are summarized in table \ref{promptresults} and on the right hand side of Fig. \ref{promptatmofluxes}.  In the same fashion as the astrophysical model tests described above, the upper limits on prompt atmospheric neutrinos were expressed in terms of the model rejection factor.  The standard calculation from Enberg \textit{et al.} which is used as the reference flux in this analysis was rejected at $90\%$ confidence level valid from an energy range between $9$ TeV and $613$ TeV.  

Under the assumptions of the present analysis, we reject  the RQPM \cite{naumov} (Recombination Quark Parton Model) at a $3\sigma$ confidence level, which strongly disfavors the authors' non perturbative quantum chromodynamics (pQCD) approach to calculating the prompt flux.  The maximum and minimum calculations from Enberg \textit{et al.} represent an allowed theoretical uncertainty band due to the authors' choices of the parton distribution function (PDF), the charm quark mass, and the factorization scale which affect the pQCD calculation of the prompt atmospheric neutrino flux.  The reference model used the MRST 2001 \cite{mrst} for the PDF, a factorization scale $\mu_{\mathrm{F}}=2m_{\mathrm{c}}$ where $m_{\mathrm{c}}$ is the charm quark mass, and a charm quark mass of $1.3$ GeV.  The theoretical uncertainty represented by the minimum and maximum calculations were obtained by varying the quark mass between $1.3$ GeV and $1.5$ GeV, varying the factorization scale between $\mu_{F}=m_{c}$ and $\mu_{F}=2m_{c}$, and varying the PDFs by using MRST 2001 or CTEQ 6 \cite{cteq6}.   We ruled out the maximum calculation at $95\%$ C.L. and the standard prediction at $90\%$ C.L. These limits favor the CTEQ 6 parameterization of the PDF, a lower quark mass, and a low factorization scale.   

\begin{table}[htp]
\caption{\label{promptresults}Upper limits on prompt atmospheric neutrinos for different models.  The upper limits are expressed in terms of the model rejection factor \cite{mrf}, which is  the percentage of the reference model rejected at the stated confidence level such that $\Phi_{\mathrm{C.L}}=\mathrm{MRF}\times\Phi_{\mathrm{ref}}$. }
\centering
\begin{tabular}{c c c c c}
\hline \hline
Model & $90\%$ & $95\%$ & $3\sigma$ & 90$\%$ Energy Range (TeV) \\ [0.5ex]
\hline
Enberg (Minimum)   & $1.25$ &$1.8$ & $3.6$ &  $9 - 615$   \\
Enberg (Standard)   & $0.73$ &$1.1$ & $2.2$ &  $9  - 613 $\\
Enberg (Maximum)  & $0.53$ &$0.85$ & $1.89$ & $9 - 610$  \\
Naumov RQPM  & $0.2$ &$0.41$ & $0.87$ & $9 - 620 $ \\
Martin MRS & $2.1$ & $4.0$ & $8.9$ & $9 - 613$ \\
\hline
\end{tabular}
\end{table}



\subsection{Measurement of the Atmospheric Neutrino Spectrum}

\begin{figure*}[htp]
\includegraphics[width=0.49\textwidth]{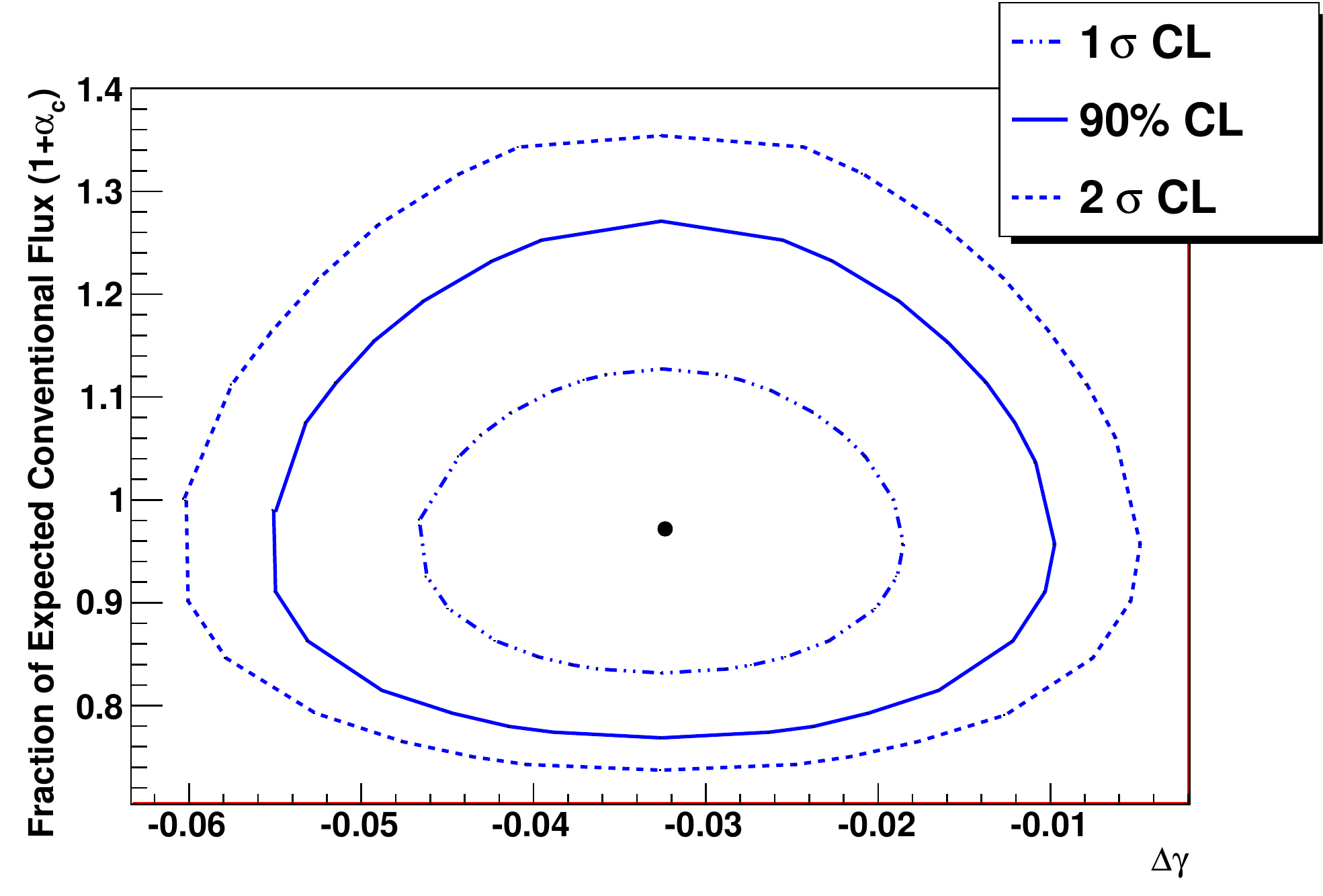}
\includegraphics[width=0.5\textwidth]{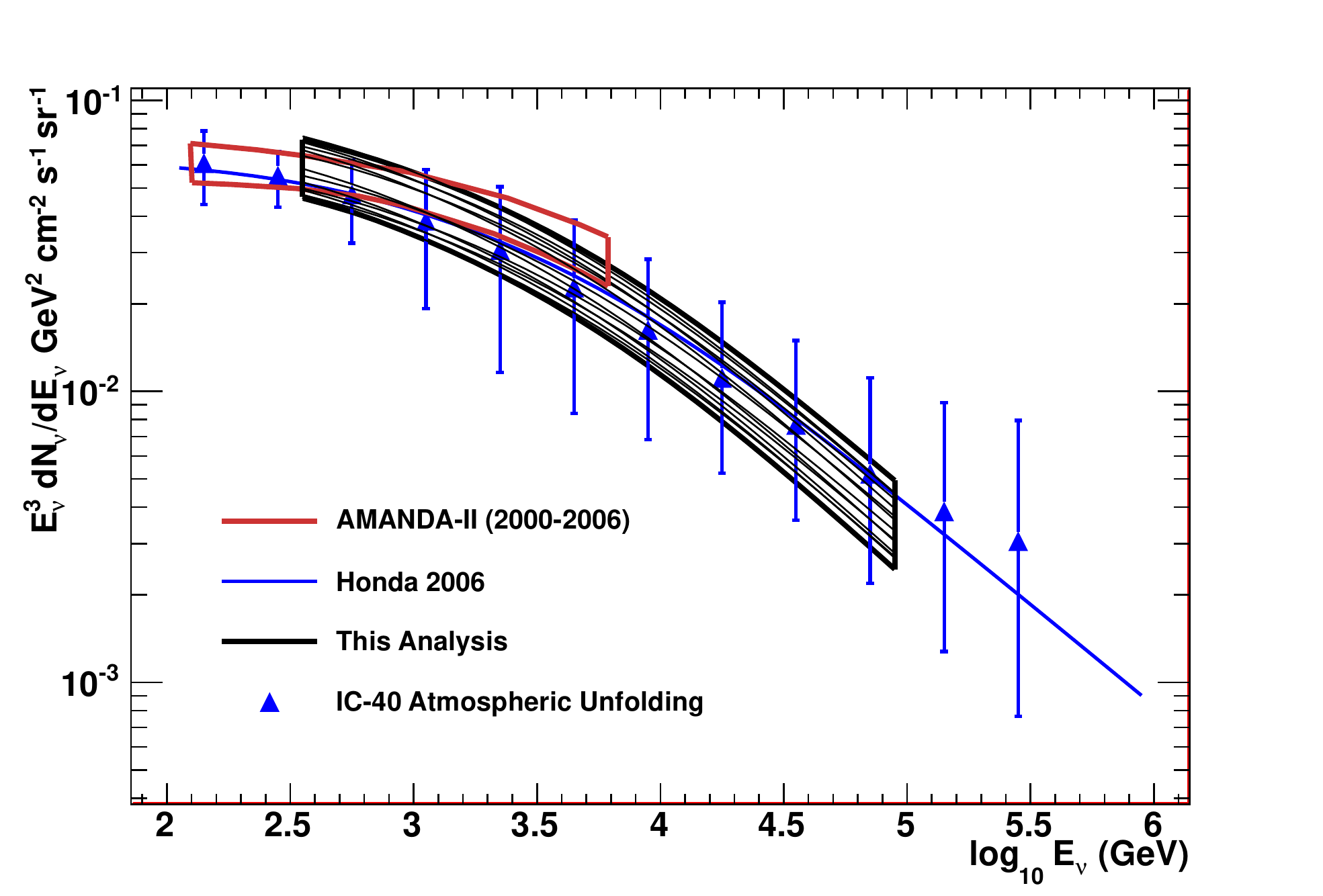}
\caption[Measured Atmospheric Neutrino Flux]{The left plot shows allowed regions for the normalization ($1+\alpha_{\mathrm{c}}$) and the change in spectral index ($\Delta\gamma$) of the conventional atmospheric neutrino flux relative to Honda \textit{et al} \cite{honda:2006}. The right plot compares the angle-averaged $\nu_{\mu} + \bar{\nu}_{\mu}$ measurement of the atmospheric neutrino flux of this work to the model prediction from Honda \textit{et al}. The fluxes are multiplied by $E^{3}$ to enhance features.  The displayed set of black curves is the band of allowed atmospheric neutrino spectra constructed from the $90\%$ boundary of the left plot in this work as described in the text.  Also shown (blue triangles) is the IceCube 40-string unfolding analysis \cite{warrenpaper}, and the AMANDA-II (empty brown band) result \cite{kelleypaper}.}  
\label{atmofluxes}
\end{figure*}

There was no evidence for astrophysical neutrinos or prompt atmospheric neutrinos in the final event sample, and therefore the final neutrino distribution was interpreted as a flux of conventional atmospheric muon neutrinos.  The profile construction method was used to measure the atmospheric neutrino flux in order to determine the normalization and any change in shape from the reference atmospheric neutrino flux model considered.  The best fit result of the atmospheric neutrino flux is of the form:
\small{
\begin{equation}
\Phi_{\mathrm{BestFit}}=(0.96\pm0.16)\left(\frac{E}{1.17 \ \mathrm{TeV}}\right)^{-0.032\pm0.014} \Phi_{\mathrm{Honda}}
\label{convphiresult}
\end{equation}
}
where the normalization of the atmospheric neutrino flux was found to be $4\%\pm16\%$ lower than the nominal prediction from Honda \textit{et al.} and the spectral index was found to be steeper by $\Delta\gamma=-0.032\pm0.014$.  The allowed regions of $(1+\alpha_{\mathrm{c}})$ and $\Delta\gamma$ are shown with the band of allowed atmospheric neutrino spectra in Fig. \ref{atmofluxes}.  The displayed band of allowed atmospheric neutrino spectra in black was constructed from the envelope of the set of curves allowed by the $90\%$ boundary in the left plot of Fig. \ref{atmofluxes}.   We note that our best fit for the spectral index rejected the nominal spectral index prediction from Honda \textit{et al.} at a $95 \%$ confidence level, neglecting the theoretical uncertainty. The overall flux normalization is consistent with the nominal model prediction at $90\%$ confidence level.  The energy range of the atmospheric neutrino flux measurement is valid from $332 \ \mathrm{GeV}$ to $84 \ \mathrm{TeV}$.  This energy range was derived from the median atmospheric neutrino energy as predicted by simulation for the lowest and highest $\mathrm{d}E_{\mathrm{reco}}/\mathrm{d}X$ values from the data. 

Also shown in the right plot of Fig. \ref{atmofluxes} is the atmospheric neutrino unfolding analysis discussed in Ref. \cite{warrenpaper} where no prior assumption was made regarding the shape of the atmospheric neutrino spectrum, while this work parametrized the atmospheric neutrino flux as a power law.  Such a bias in the prior assumption of the atmospheric neutrino flux resulted in tighter error bars than in the unfolding analysis.

\section{Conclusions and Outlook}

To summarize, we have set the field's most stringent limit on astrophysical muon neutrinos from unresolved sources.  The $90\%$ upper limit on an  astrophysical flux with an $E^{-2}$ spectrum is $8.9 \times 10^{-9} \ \mathrm{GeV \ cm^{-2} \ s^{-1} \ sr^{-1}}$, valid from the energy range of $35$ TeV to $7$ PeV.  Several optimistic astrophysical neutrino production models have been rejected at a $5\sigma$ confidence level.  We have set limits on the prompt component of the atmospheric neutrino flux, with a preference for perturbative QCD models in the energy range between $9$ TeV and $613$ TeV.  Finally, we have also measured the atmospheric muon neutrino flux from $332$ GeV to $84$  TeV and found a fit to the overall normalization of the atmospheric neutrino flux that is $4\%$ lower than the calculation from Honda \textit{et al.} \cite{honda:2006}.  The preferred spectrum is somewhat steeper than the assumed extrapolation of the Honda spectrum, perhaps reflecting the steepening of the spectrum associated with the knee, as discussed in Ref. \cite{ILLM}.  Overall, our result here is consistent with other measurements made of the atmospheric neutrino flux with the IceCube detector. The $90\%$ error band on the measured flux overlaps with the $90\%$ error band of the result from IceCube's predecessor, the AMANDA-II experiment \cite{kelleypaper}.  

The $90\%$ upper limits on the astrophysical $\nu_{\mu}$ models and the prompt component of the atmospheric neutrino flux are dependent on the assumptions made on the conventional atmospheric neutrino flux.   We extrapolated the model from Honda \textit{et al.} above the maximum calculated energy of 10 TeV by assuming the conventional spectrum continued to follow its approximate $E^{-3.7}$ spectral shape. This extrapolation is dependent on the location and primary cosmic-ray composition of the knee which are currently not well known.  The spectrum of all nuclei steepens in an energy region around 3 PeV total energy per nucleus.  With standard assumptions about the composition, and assuming that the underlying physics depends on magnetic rigidity, the spectrum of protons must become steeper around 1 PeV or lower.  This has the consequence that the spectrum of neutrinos from decay of pions and kaons must steepen at around 100 TeV \cite{ILLM}, which is just in the crossover region for the prompt component.  What is needed is a detailed calculation of atmospheric neutrinos based on a realistic treatment of the primary cosmic-ray spectrum and composition through the knee region that extends beyond the current limits of 10 TeV.  Because of the steepening, the limits on prompt neutrinos in Fig. \ref{promptatmofluxes} will be relaxed to some extent.  For future work, it is critically important also to obtain more precise measurements of the primary composition and spectra in the knee region.  The KASCADE experiment~\cite{Antoni} already suggests that the proton component is suppressed in the knee region.  IceCube itself, with its surface component IceTop, has the potential to measure the spectrum and composition in the knee region and beyond.


The stringent $90\%$ upper limit on a diffuse astrophysical flux of muon neutrinos reported by this work implies that the IceCube detector in its 40 string configuration (as used in this analysis) is not yet sensitive enough to discover astrophysical neutrinos from unresolved sources. The full 86-string array has been completed during the 2010-2011 summer construction season at the South Pole.  A $5\sigma$ discovery of an astrophysical $E^{-2} \ \nu_{\mu}$ flux at the $90\%$ limit derived by this work will take three years of the full IceCube array.  

This time scale for discovery can be made shorter by an improved understanding of the various sources of systematic uncertainty and considering new analysis techniques. With a proper measurement of the prompt component of the atmospheric neutrino flux, the time scale for discovery becomes more tractable.   In this analysis we have not yet made use of the difference in angular behavior of prompt neutrinos (which are isotropic) and conventional atmospheric neutrinos (which have a higher intensity near the horizon).  Analyses dedicated to the study of leptons from the decay of charmed mesons would also yield a better understanding of the physics of air showers and atmospheric neutrinos. Strategies other than using atmospheric $\nu_{\mu}$ to search for the prompt component involve a thorough investigation of the down-going muon flux and a measurement of the atmospheric neutrino spectrum from $\nu_{\mathrm{e}}$.   The measurement of the atmospheric $\nu_{\mathrm{e}}$ flux has an advantage that the transition energy from conventional $\nu_{\mathrm{e}}$ to prompt $\nu_{\mathrm{e}}$ occurs at an order of magnitude lower in energy than in $\nu_{\mu}$.

The event selection in this analysis used the Earth as a filter to remove the large down-going atmospheric muon background.  An improved simulation of atmospheric muons would allow a diffuse astrophysical $\nu_{\mu}$ search to incorporate the down-going region in the analysis and search for astrophysical neutrinos over the entire sky. We note that there is a slight tilt in the measured angular distribution of atmospheric neutrinos with respect to our extrapolation based on the angular dependence in ~\cite{honda:2006}, as illustrated in Fig. \ref{coszen}. This  discrepancy did not affect our limit on astrophysical $\nu_{\mu}$ or our reconstructed atmospheric neutrino spectrum. However, understanding the origin of the discrepancy is important for future work. 

Although this analysis focuses on $\nu_{\mu}$, IceCube is sensitive to all flavors of neutrinos.  As the detector grows, reconstruction methods mature, and the understanding of the various sources of systematic uncertainty improve, it would be possible to combine event topologies from different neutrino flavors in a multi-flavor analysis.  A simultaneous search for neutrinos of all flavors from unresolved astrophysical sources would be significantly more sensitive than an analysis focusing exclusively on a single neutrino flavor.  

\begin{acknowledgments}

We acknowledge the support from the following agencies:
U.S. National Science Foundation-Office of Polar Programs,
U.S. National Science Foundation-Physics Division,
University of Wisconsin Alumni Research Foundation,
the Grid Laboratory Of Wisconsin (GLOW) grid infrastructure at the University of Wisconsin - Madison, the Open Science Grid (OSG) grid infrastructure;
U.S. Department of Energy, and National Energy Research Scientific Computing Center,
the Louisiana Optical Network Initiative (LONI) grid computing resources;
National Science and Engineering Research Council of Canada;
Swedish Research Council,
Swedish Polar Research Secretariat,
Swedish National Infrastructure for Computing (SNIC),
and Knut and Alice Wallenberg Foundation, Sweden;
German Ministry for Education and Research (BMBF),
Deutsche Forschungsgemeinschaft (DFG),
Research Department of Plasmas with Complex Interactions (Bochum), Germany;
Fund for Scientific Research (FNRS-FWO),
FWO Odysseus programme,
Flanders Institute to encourage scientific and technological research in industry (IWT),
Belgian Federal Science Policy Office (Belspo);
University of Oxford, United Kingdom;
Marsden Fund, New Zealand;
Japan Society for Promotion of Science (JSPS);
the Swiss National Science Foundation (SNSF), Switzerland;
A.~Gro{\ss} acknowledges support by the EU Marie Curie OIF Program;
J.~P.~Rodrigues acknowledges support by the Capes Foundation, Ministry of Education of Brazil.

\end{acknowledgments}

%

\end{document}